\documentclass[manuscript]{acmart}
\AtBeginDocument{%
  }

\setcopyright{acmlicensed}
\copyrightyear{2025}
\acmYear{2025}
\acmDOI{XXXXXXX.XXXXXXX}

\acmConference[Conference acronym 'XX]{Make sure to enter the correct
  conference title from your rights confirmation email}{June 03--05,
  2018}{Woodstock, NY}
\usepackage[colorinlistoftodos]{todonotes}
\presetkeys{todonotes}{color=orange!80}{}

\usepackage{pgfplots}
\pgfplotsset{compat=newest}
\usepackage{pgfplotstable}
\usetikzlibrary{matrix, positioning}
\usepgfplotslibrary{groupplots}

\definecolor{skyblue}{RGB}{75,206,235}
\definecolor{orange}{RGB}{255,165,0}
\usepackage{multirow}
\usepackage[table,xcdraw]{xcolor}
\usepackage[dvipsnames]{xcolor}
\usepackage{array}
\usepackage{colortbl}
\usepackage{makecell}
\usepackage{longtable}
\usepackage{tikz}
\usetikzlibrary{shapes.geometric, arrows.meta, positioning}
\usetikzlibrary{shapes.misc, backgrounds, positioning, calc} 
\usepackage{booktabs}
\usepackage{adjustbox}
\usepackage{caption}
\usepackage{soul}
\usepackage{float} % for [H] placement
\usepackage{tabularx}
\usepackage[acronym]{glossaries}
\glsdisablehyper
\newacronym{AI}{AI}{Artificial Intelligence}
\newacronym{SE}{SE}{software engineering}
\newacronym{HCI}{HCI}{human-computer interaction}
\newacronym{XAI}{XAI}{explainable AI}
\newacronym{LLMs}{LLMs}{large language models}
\newacronym{HAI}{HAI}{human-AI interaction}
\newacronym{CS}{CS}{computer science}
\newacronym{LMS}{LMS}{Library Management System}
\newacronym{E-comm}{E-comm}{E-commerce}
\newacronym{IDE}{IDE}{integrated development environment}

\usepackage{longtable}
\usepackage{ltablex}
\usepackage{xspace}
\keepXColumns

\newcommand{\RQ}[1]{\textbf{RQ\textsubscript{#1}}}

\newcommand{\deltacell}[1]{%
  \begin{tikzpicture}
    \fill[black!70] (0,0) rectangle (#1/50,0.3);
    \node[anchor=mid west, font=\small] at (#1/50 + 0.1, 0.15) {#1\%};
  \end{tikzpicture}%
}

\newcommand{\boxwidth}{0.12\textwidth}
\newcommand{\boxheight}{1.2cm}

\tikzstyle{Briefing} = [rectangle, minimum width=\boxwidth, minimum height=\boxheight, text centered, draw=black, fill=green!30]
\tikzstyle{PreExpSurvey} = [rectangle, minimum width=\boxwidth, minimum height=\boxheight, text centered, draw=black, fill=blue!30]
\tikzstyle{Calibration} = [rectangle, minimum width=\boxwidth, minimum height=\boxheight, text centered, draw=black, fill=orange!30]
\tikzstyle{Walkthrough} = [rectangle, minimum width=\boxwidth, minimum height=\boxheight, text centered, draw=black, fill=purple!30]
\tikzstyle{Experiment} = [rectangle, minimum width=\boxwidth, minimum height=\boxheight, text centered, draw=black, fill=red!30]
\tikzstyle{Interview} = [rectangle, minimum width=\boxwidth, minimum height=\boxheight, text centered, draw=black, fill=magenta!30]
\tikzstyle{Debriefing} = [rectangle, minimum width=\boxwidth, minimum height=\boxheight, text centered, draw=black, fill=gray!30]
\tikzstyle{arrow} = [thick,->,>=stealth]

\begin{document}

%%
%% The "title" command has an optional parameter,
%% allowing the author to define a "short title" to be used in page headers.
\title{GazePrinter: Visualizing Expert Gaze to Guide Novices in a New Codebase}

%%
%% The "author" command and its associated commands are used to define
%% the authors and their affiliations.
%% Of note is the shared affiliation of the first two authors, and the
%% "authornote" and "authornotemark" commands
%% used to denote shared contribution to the research.
\author{Peng Kuang}
% \authornote{Both authors contributed equally to this research.}
\email{peng.kuang@cs.lth.se}
\orcid{0000-0002-7029-5655}
\affiliation{%
  \institution{Lund University}
  \city{Lund}
  \state{Skåne}
  \country{Sweden}
}

\author{Emma S{\"o}derberg}
% \authornotemark[1]
\email{emma.soderberg@cs.lth.se}
\orcid{0000-0001-7966-4560}
\affiliation{%
  \institution{Lund University}
  \city{Lund}
  \state{Skåne}
  \country{Sweden}
}

\author{April Yi Wang}
\affiliation{%
  \institution{ETH Z{\"u}rich}
  \city{Z{\"u}rich}
  \state{Z{\"u}rich}
  \country{Switzerland}}
\email{april.wang@inf.ethz.ch}
\orcid{0000-0001-8724-4662}

%\author{Diederick C. Niehorster}
%\orcid{0000-0002-4672-8756}
%\affiliation{%
%  \institution{Humanities Lab \& Dept. of Psychology, Lund University}
%  \city{Lund}
%  \country{Sweden}
%}
%\email{diederick_c.niehorster@humlab.lu.se}

\author{Martin Höst}
\affiliation{%
  \institution{Malmö University}
  \city{Malmö}
  \country{Sweden}}
\email{martin.host@mau.se}
\orcid{0000-0002-9360-8693}

%%
%% By default, the full list of authors will be used in the page
%% headers. Often, this list is too long, and will overlap
%% other information printed in the page headers. This command allows
%% the author to define a more concise list
%% of authors' names for this purpose.
\renewcommand{\shortauthors}{Kuang et al.}

%%
%% The abstract is a short summary of the work to be presented in the
%% article.
\begin{abstract}
Program comprehension is an essential activity in software engineering. Not only does it often challenge professionals, but it can also hinder novices from advancing their programming skills. Gaze, an emerging modality in developer tools, has so far primarily been utilized to improve our understanding of programmers' visual attention and as a means to reason about programmers' cognitive processes. There has been limited exploration of integrating gaze-based assistance into development environments to support programmers, despite the tight links between attention and gaze. We also know that joint attention is important in collaboration, further suggesting that there is value in exploring collective gaze.

In this paper, we investigate the effect of visualizing gaze patterns gathered from experts to novice programmers to assist them with program comprehension in a new codebase. To this end, we present GazePrinter, designed to provide gaze-orienting visual cues informed by experts to aid novices with program comprehension. We present the results of a mixed-methods study conducted with 40 novices to study the effects of using GazePrinter for program comprehension tasks. The study included a survey, a controlled experiment, and interviews. 
We found that visualization of expert gaze can have a significant effect on novice programmers' behavior in terms of which path they take through the code base; with GazePrinter, novices took a path closer to the path taken by experts. We also found indications of reduced time and cognitive load among novices using GazePrinter. 
%
%Our study contributes to bridging the gap between program comprehension research, gaze prototyping studies, and controlled experiments with programmers, integrating both quantitative and qualitative perspectives. 

% No need to mention future work in the abstract.
% Future work can continue to investigate how non-professional programmers formulate and execute their strategies for comprehending code mixed with natural text prompts at scale and within AI-native programming environments. 
\end{abstract}

%%
%% The code below is generated by the tool at http://dl.acm.org/ccs.cfm.
%% Please copy and paste the code instead of the example below.
%%

%%
%% Keywords. The author(s) should pick words that accurately describe
%% the work being presented. Separate the keywords with commas.

\received{12 Mar 2026}
% \received[revised]{12 June 2026}
% \received[accepted]{30 June 2026}

%%
%% This command processes the author and affiliation and title
%% information and builds the first part of the formatted document.
\maketitle

\section{Introduction}
\label{sec:intro}

Program comprehension is a crucial task recognized by both academia and industry in \gls{SE}. Among its variants, ranging from small single-file program to programs stretching over many files, comprehension of a program stretching over a codebase may be the most challenging due to scale and complexity. 
Comprehending programs of this size is not only a challenge for professional developers~\cite{kuang2024devsurvey, zhao2024earlycareer, green2023onboarding}, but also non-professional programmers such as computer science students working on group projects and scientists who need to reuse peers' code for data analysis. In order to learn, reuse, or deploy software, programmers need to understand at least part of (if not the entirety of) a codebase. 

Understanding a program of the scale of a codebase adds additional challenges.
The code within a shared codebase is usually co-authored and maintained by quite a few programmers from different roles, teams, and locations, as well as at different points in time. This often introduces different code styles and ways of thinking, making it inherently difficult to read and understand, especially for the first time. This situation is exacerbated as the volume of code increases. Consequently, approaching a new codebase can be time-consuming~\cite{green2023onboarding}, mentally demanding~\cite{gopstein2017atom, gopstein2018confusion}, and overwhelming~\cite{bexell2024firstbug}. There is a need for better tool support for program comprehension, especially as the size of the code increases.
Researchers and practitioners have been developing a suite of techniques to assist in program comprehension. For example, linters (e.g. CheckStyle~\cite{checkstyle2025checkstyle}) to synchronize style, shared documentation formats (e.g. JavaDoc~\cite{wikipedia2025javadoc}) to support understandability, visualization~\cite{cornelissen2010experiment, hawes2015codesurveyor} to support comprehension, and walkthroughs~\cite{taylor2022codetour} to guide developers through a codebase. Despite the broad adoption of these tools, many of the challenges with program comprehension remain, especially for large programs, suggesting that this is still an avenue worthy of more exploration. 

An idea investigated in this work is to use gaze data from experienced developers to guide developers new to a codebase in their comprehension of the codebase. 
Program comprehension is closely related to code reading, a task that has previously been studied using gaze analysis. However, exploration of gaze-orienting developer assistance, where gaze data gathered by an eye-tracker and analyzed as an integrated part of a developer tool, is still limited.
Gaze-orienting assistance has the potential to capture subjects' attention and to inform assistance via gaze pattern analysis. For example, Saranpää et al.~\cite{saranpaa2023gander} present the GANDER platform with a proof-of-concept code review assistant showing the relationships between variable declarations and uses based on detected gaze patterns, and Santos et al.~\cite{santos2021javardeye} present the Javardeye code editor supporting gaze-driven code selection and scrolling. These are examples of gaze-orienting assistance, in which gaze is analyzed during the interaction between one user and the tool. 

Another avenue of exploration is visualization aimed at guiding the user's gaze. Bednarik et al.~\cite{bednarik2018emme} present a classroom experiment with 30 students and two single-file programs from the EMIP dataset, using an intervention in which a predefined eye movement model, defined by the program comprehension theory, guides the attention of participants during comprehension tasks. Cheng et al~\cite{cheng2022collaborative} present an experiment with 39 participants and six small single-file programs, using an intervention in which attention is guided by the gaze of other users during code review tasks. In both cases, they find positive effects of the intervention. We see a possibility to expand on these design ideas towards an approach where expert gaze gathered during a program comprehension task is encoded in a design to guide the attention of novices during the same program comprehension task. That is, use an eye movement model, similar to Bednarik et al., but informed by the gaze of others, similar to Cheng et al., specifically the gaze gathered from experts motivated by the positive effects of cognitive apprenticeship~\cite{collins2015cognitive}. We further see a possibility of going beyond program comprehension of single files to program comprehension of codebases, a task commonly facing novices as they enter the industry.

In this paper, we present a study aimed at addressing the following research questions (RQ):
\begin{itemize}
    \item[\RQ{1~~}] \textbf{To what extent can visualization of expert gaze assist novices in program comprehension of a new codebase?} 
    We focus on comprehension of programs on the scale and complexity of a codebase and on novices, as we see support for comprehension of larger programs as an area in need of more tool support, especially for novices who have less experience. 
    \item[\RQ{2}] \textbf{To what extent do visualization of expert gaze facilitate the immediate transfer of reading strategies for novices in program comprehension of a new codebase?} Given that it may be challenging to gather expert gaze for a task, it is interesting to consider if there are learning effects with regard to reading strategies between similar program comprehension tasks.
    \item[\RQ{3}] \textbf{To what extent do visualization of expert gaze influence novices' learning experience in relation to software engineering?}~
    % The presence of expert gaze can implicitly create a sense of support. The experience of being supported can impact novices' confidence in and perception of software engineering.
    % It may be good to be a bit more general here and steer away a bit more from assumed improvement
    Visualization of expert gaze provides support in a judgment-free and non-pressing way. Receiving such support for a task could potentially have a positive effect on learning~\cite{happe2021multidisciplinarity} among novices, for instance, by building of more confidence, enabling a better learning experience. 
    \item[\RQ{4}] \textbf{What is the user experience of visualization of expert gaze during program comprehension of a new codebase?} In addition to comprehension performance and learning, we deem that user experience also reflects the quality of the assistance. The user experience shapes programmers' perception of the assistance, which is an important factor for them to adopt or abandon it. 
\end{itemize}

To this end, we developed a gaze-orienting design, where gaze behavior gathered from experts during program comprehension tasks on a codebase is visualized to aid novices in program comprehension on the same codebase, inspired by the ideas from Bednarik et al.~\cite{bednarik2018emme} as well as built on our previous work~\cite{kuang2024designing, kuang2024devsurvey}. We realized this design in a prototype called GazePrinter, which is integrated as a plugin in the main stream \gls{IDE} IntelliJ. We then designed a mixed-methods study incorporating a survey, a controlled experiment, where GazePrinter is used as the treatment to represent assistance through visualization of expert gaze, and interviews.
The controlled experiment includes tasks with code reading and code editing in two codebases of intermediate size and complexity, placing them towards the middle in the range between simple code snippets and large-scale codebases that have been maintained for many years. The study was conducted with a relatively homogeneous group of 40 novice programmers.

The contributions of this paper are the following: 
\begin{itemize}
\item a design for visualization of expert gaze to guide attention during code reading, 
\item a prototype, GazePrinter, realizing our design for the visualization of expert gaze, and
\item an evaluation with a mixed-methods study, including a controlled experiment where we find significant quantitative effects on reading behavior. That is, the experiment group reads files in an order statistically distinct from the control group when our tool is present. 
\end{itemize}

%to gather quantitative data in addition to questionnaires and interviews, enabling a more systematic and deeper understanding. 

The remainder of this paper is organized as follows. Section~\ref{sec:background} provides background and a review of related work. Section~\ref{sec:gaze-printer} presents the design and implementation of GazePrinter. Section~\ref{sec:method} describes our study in more detail. Section~\ref{sec:results} presents the results. Section~\ref{sec:discussion} addresses the research questions and discusses the implications and future work. Finally, Section~\ref{sec:conclusion} concludes the paper. 
\section{Background and Related Work}
\label{sec:background}

\subsection{Program Comprehension}

In this section, we provide a brief review of the literature with respect to the definition of program comprehension and strategies for program comprehension.

% \todo[author=Peng]{write about program comprehension literature and define different strategies used in this paper}
\subsubsection{Definition of Program Comprehension}

Program comprehension as a term has evolved over time and there are different definitions. Program comprehension may also be called code comprehension, source code comprehension, or codebase comprehension. The specific wording in some cases is an indication of size of the program or code in question, ranging from code snippets in one file to full codebases with many files.
Wyrich~\cite{wyrich2023program} defines source code comprehension as \emph{``a person’s intentional act and degree of accomplishment in inferring the meaning of source code''}. We adopt this definition (code comprehension for short) and deem that it can be used interchangeably with program comprehension if their differentiation is not explicitly indicated by researchers. However, program comprehension can also refer to a holistic view of a software project that also covers software architecture, documentation, diagrams, and other materials in addition to source code, while code comprehension ideally only deals with code (but sometimes researchers may also include related code comments when reporting the study). In this context, we view code comprehension as a subtype or subset of program comprehension, similar to the perspective of Wyrich~\cite{wyrich2023designexp}. We further see a need to be explicit about the program under study with regard to size and shape:

\begin{itemize}

\item \textbf{Size of the Program}.
We need to consider the scale of the program under study; a program can range from a single file to a large codebase with a large number of files. We believe that it is reasonable to assume that size has an effect on how a program is comprehended. For example, we may observe different behavioral patterns between an eye-tracking study using a self-contained single-file program and one using a program that includes multiple files. 

\item \textbf{Shape of the Program}. 
We need to consider the shape of the program under study; a codebase with a program may include many non-code files (e.g., documentation, requirements, user manuals), and for files with code, there may be use of several different programming languages. Documentation may be included in a code base but may also be external (e.g., documentation of APIs and libraries). Especially in the case where the material under study is a code base, we believe it is reasonable to assume that the shape of the code base has an effect on how the program is comprehended. For example, a participant may wander between documentation files and code to gain an understanding of how the program works.

%We also see a need to draw a boundary with program comprehension if it also contains, in particular, documentation and other assistant, non-code materials. Whether or not these materials reside within the same repository or codebase as the source code should be specified. This is because, for instance, documentation is also a broad term that encompasses developer comments, developer guides (e.g., the readme.md file), user manuals, requirement specifications, and external documentation for certain APIs or libraries used in the source code. 

\end{itemize}

In our study, we use a program in the form of a code base with many files, both code and non-code, but with one programming language for code files (Java). We consider the size of the code base to be medium-sized, that is, it is not on the scale of an industrial codebase, but much larger than a short, single-file program.

%In this study, we use the term `codebase comprehension' to indicate that the stimulus studied is a comprehensive program that contains mostly source code but also documentation, which typically accompanies the code in a software repository. The size of the codebase is medium-sized. It is not the scale comparable to industrial codebases, but it is also not a short, single-file program.

\subsubsection{Program Comprehension Strategies}

During the past 40 years, there have been several attempts to describe and classify approaches to program comprehension. Due to a variation in approaches and on which level of abstraction the program comprehension task has been considered, this work has resulted in several documented strategies, models, and measures related to program comprehension. 
We list commonly reported strategies~\cite{harth2017program,siegmund2016program,storey2005program} in Table~\ref{tab:strategies} and describe them briefly below:

%\textbf{top-down}, \textbf{bottom-up}, \textbf{hybrid}, \textbf{as-needed vs. systematic}, and \textbf{integrated}. 

%
%In the following, we briefly describe what each program comprehension strategy/model means based on Harth and Dugerdil's synthesis~\cite{harth2017program} and Storey's summarization~\cite{storey2005program}. 

\begin{itemize}

\item \textbf{Bottom-up}.
When using this strategy, the programmer approaches the code from the low-level constructs to build a global overview of what the software does. The programmer tends to read the code statement by statement and then group what they have read into more abstract information (chunking). By grouping these chunks recursively into higher-level abstractions, the programmer eventually achieves a global understanding of the program. 
% The concepts of syntactic and semantic knowledge of programs, introduced by Shneiderman and Mayer~\cite{shneiderman1979semantic, storey2005program}, are relevant to this strategy. Syntactic knowledge is language-dependent and concerns the syntactic units in a program, while semantic knowledge is language-independent and is constructed progressively into a multi-level mental representation that makes sense to the programmer.
% The concepts of program model and situation model proposed by Pennington~\cite{pennington1987situation}, are also relevant to this strategy. The program model concerns the control flow, and the situation model concerns the data flow and functional connections. The program model is believed to precede the situation model in the comprehension process. 

\item \textbf{Top-down}.
When using this strategy, the programmer will formulate some hypotheses about the program and then try to map these to the code. It is usually observed when the programmer has some prior knowledge about the code. 
% This strategy is adapted from the ``knowledge-based strategy'' by Exton~\cite{exton2002constructivism}. 
%
% O'Brien et. al~\cite{obrien2004program} further proposed that expectation-based and inference-based comprehension processes should be distinguished in the traditionally top-down strategy. 

\item \textbf{Systematic}.
%This strategy describes two fractions of approaches that programmers may apply according to their needs or goals. 
A systematic strategy is deemed to lead to a more thorough understanding of the program because it creates not only static knowledge about the structure of the code but also causal knowledge about how its components interact with each other. 

\item \textbf{As-needed}.
An as-needed strategy is for sufficient understanding at the time. It often is a partial understanding of the program, since the programmer only reads the code closely relevant to their goal, e.g., a task.

\end{itemize}

%\item \textbf{Integrated/Hybrid}.
In addition to the above strategies, there are \textbf{integrated} or \textbf{hybrid} variants. The integrated strategy, originally called the ``integrated metamodel'' by von Mayrhauser and Vans~\cite{mayrhauser1993integrated}, mixes multiple previous strategies. A programmer may adopt any of the strategies mentioned above in the process of understanding a program. The top-down model may manifest if the programmer has prior knowledge in relation to the code; as such, the programmer may formulate hypotheses when starting to read. When the programmer encounters code completely new, the program and situation models may be activated. In the process of building the situation model, the programmer may adopt either an opportunistic (as-needed) or systematic strategy. 
During comprehension, the programmer keeps adding new and inferred information pieces to the knowledge cluster they keep in mind for the program, which also interacts with their existing knowledge base. 
Additionally, there is a hybrid strategy~\cite{harth2017program}, which is the combination of top-down and bottom-up. We interpret this strategy as a simpler subvariant of the integrated metamodel.
%, as we treat it as a simple version of it. 

% \input{figs/strategy-relationship}

\begin{figure}[th]
\centering
\begin{adjustbox}{center}
\scalebox{0.75}{
\begin{tikzpicture}

  % Circle A - Integrated Metamodel
  \fill[cyan!10] (0.1,0) circle (5cm);
  \draw[thick, cyan] (0.1,0) circle (5cm);
  \node[cyan!80!black, font=\bfseries] at (0.1,5.5) {Integrated Strategy};

  % Circle B - As-needed
  \fill[green!20, opacity=0.7] (-2,1.5) circle (2cm);
  \draw[thick, green!60!black] (-2,1.5) circle (2cm);
  \node[green!60!black, font=\bfseries] at (-2,1.5) {As-needed};

  % Circle C - Systematic
  \fill[orange!20, opacity=0.7] (1.1,-1.2) circle (3.2cm);
  \draw[thick, orange!80!black] (1.1,-1.2) circle (3.2cm);
  \node[orange!80!black, font=\bfseries] at (1.2, 2.3) {Systematic};

    % Circle H - Hybrid
  \fill[cyan!20, opacity=0.5] (1.1,-1.2) circle (3cm);
  \draw[thick, cyan!80!black] (1.1,-1.2) circle (3cm);
  \node[cyan!80!black, font=\bfseries] at (1.1, -4) {Hybrid};

  % Circle D - Top-down
  \fill[red!20] (0,-2.2) circle (1.4cm);
  \draw[thick, red!80!black] (0,-2.2) circle (1.4cm);
  \node[red!80!black, font=\bfseries] at (0,-2.2) {Top-down};

  % Circle E - Bottom-up
  \fill[violet!20] (2.4,-0.5) circle (1.4cm);
  \draw[thick, violet!80!black] (2.4,-0.5) circle (1.4cm);
  \node[violet!80!black, font=\bfseries] at (2.4,-0.5) {Bottom-up};

\end{tikzpicture}
}
\end{adjustbox}
\caption{Illustration of how we relate program comprehension strategies to each other.}
\label{bkg:strategy-relation}
\end{figure}
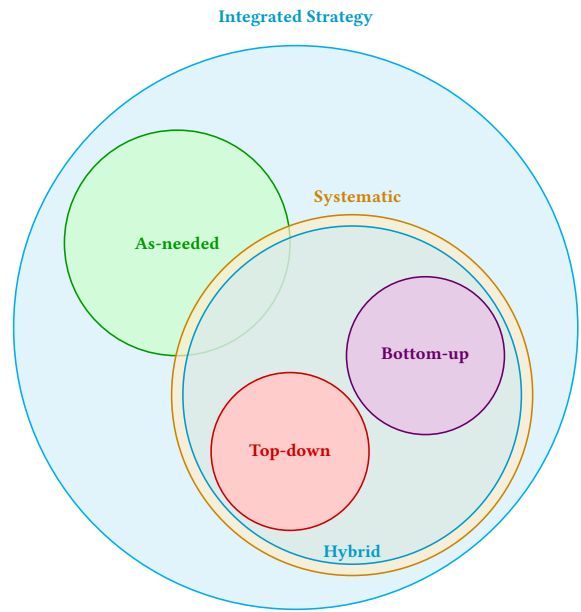

Figure~\ref{bkg:strategy-relation} provides an overview of how we relate these strategies to each other.
 Although these models capture the common patterns that exist among different groups of programmers and provide good ground for explaining programmers' reading behaviors and cognitive processes in understanding a program, it is worth discussing their weaknesses. 
 First, as exemplified by the last metamodel, they are not mutually exclusive. For example, the systematic approach may be an exhaustive bottom-up approach or an experienced top-down strategy. It can even encompass the as-needed strategy for a functionality of the program if it is written in a verbose manner. And if it is a hybrid of any of them, it can be viewed as a naive version of the integrated metamodel. Second, the context is not always clear. Although some of these models mentioned the premise that the program is familiar or unfamiliar to the programmer, some did not take this into account. This is an important factor because even for the same program, a programmer may use a bottom-up or as-needed approach at the first round but a top-down or systematic approach at later rounds. Third, the scale of the program may also influence how the programmer approaches it. For instance, with a short program, the programmer may use a bottom-up approach regardless. However, for a large program, the programmer may be inclined to apply the integrated model or an as-needed strategy.

\begin{table}[ht]
\caption{Overview of papers providing a review of program comprehension strategies.}
\label{tab:strategies}
\centering
\begin{tabular}{lllllll}
\hline
\textbf{Paper} & \textbf{Bottom-up} & \textbf{Top-down} & \textbf{Systematic} & \textbf{As-needed} & \textbf{Integrated} & \textbf{Hybrid} \\
\hline
(Storey, 2005)~\cite{storey2005program} &x &x &x &x &x & \\ 
(Siegmund, 2016)~\cite{siegmund2016program} &x &x & & &x & \\
(Harth \& Dugerdil, 2017)~\cite{harth2017program} &x &x &x &x &x &x \\
\hline
\end{tabular}
\end{table}

%We believe that strategy is something the subject may consciously formulate beforehand. It corresponds more to the top-down model. In our view, the bottom-up model is more of a pattern that the programmer unconsciously exhibits. We thus refer to it as an approach to indicate this nuanced difference. 

\subsubsection{Design for Program Comprehension}
%codetour, codecompass,
We present a review of designs aiming to assist with program comprehension of larger programs in the shape of a codebase.
Begel et al.~\cite{begel2010codebook} proposed and implemented a framework, CodeBook, to help developers within the same organization to more easily reach domain experts for inter-team collaboration. Their design visualized the relationship between code artifacts and their owners to pinpoint key points of contact. They evaluated their designs on several codebases with a total of 19 software engineers and testers and received positive feedback. 
Hawes et al.~\cite{hawes2015codesurveyor} present an algorithm designed to visualize the Linux kernel codebase of 1.4 MLOC (million lines of code) into continents, lakes, and bays. They conducted a small qualitative study with five programmers to evaluate the prototype and found some analogical and navigational gaps between programmers' expectations and their assumptions. 
Taylor and Clarke~\cite{taylor2022codetour} conducted a usability study for an onboarding tool, CodeTour, with fifteen professional developers on a Java project with about 600 LOC. Their evaluation suggested that the themed textual annotations provided by CodeTour could cue developers to locate relevant code more easily and avoid wasting effort checking irrelevant files. 
Balfroid et al.~\cite{balfroid2024llmonboarding} examined the feasibility of using LLMs to generate code tours to onboard new developers. They found that the explanations generated by the model lacked prioritization, links, and context. It further produced repetitive information already covered by comments and method signatures. These studies primarily targeted professional scenarios and collected mostly preliminary qualitative results (if included).

\subsection{Gaze-based Assistance in Software Engineering} 

% gaze-based tool design or prototype
We discuss the software engineering (SE) studies that employ gaze as a type of assistance or control in a prototype. A line of research leveraged gaze to assist programmer collaboration, particularly code maintenance and review. Ahren et al.~\cite{ahrens2019eyetracking} investigated whether gaze-based attention information can help programmers navigate and locate relevant code files by incorporating coarse-grained heatmaps and class name coloring into the IDE. Hijazi et al.~\cite{hijazi2021ireview} prototyped the iReview system to predict the quality of code review based on programmers' gaze and heart rate variability, aiming to enhance software reliability. 
Cheng et al. \cite{cheng2022collaborative} developed a prototype that visualizes connections between the code under review and its related source code. They reported that such assistance improved programmers' code review efficiency. 
Saranpää et al.~\cite{saranpaa2023gander} developed the GANDER platform for exploration of gaze-based assistance in code review and study of gaze behavior during code review. The platform supports dynamic gaze tracing and replay in their system. Another line of research focuses on employing gaze as a driver for source code navigation and selection, including EyeDE~\cite{glucker14eyede}, EyeNav~\cite{radevski2016eyenav}, CodeGazer~\cite{shakil2019codegazer}, JavardEye~\cite{santos2021javardeye}. Their primary finding is that programmers may favor gaze-based navigation for variable or method declarations jumping. 

Overall, most of these studies (except~\cite{ahrens2019eyetracking}) tapped into real-time gaze and concentrated on synchronous collaboration and interactions. The work of Ahren et al.~\cite{ahrens2019eyetracking} had merits in utilizing a small program with multiple files and class name coloring. However, their gaze-based assistance design lacked precision and color fine-tuning, which they reported had affected the code clarity and readability.

\subsection{Controlled Experiment with Programmers} 

We discuss the controlled experiment studies in SE conducted with programmers on the topic of program comprehension. 
Cornelissen et al.~\cite{cornelissen2010experiment} experimented with 34 non-novice programmers to examine whether an execution trace visualization tool can facilitate program comprehension. They found a significant improvement in time spent and task correctness comprehending the CHECKSTYLE project (written in Java, 59 KLOC (thousand lines of code)) among participants who had access to their tool. Jbara et al.~\cite{jbara2017experiment} experimented with twenty programmers to investigate how programmers read C programs that encompass repetitive code constructs. They found that programmers tend to skim code with a recurring pattern and read in a non-linear manner. 
Park et al.~\cite{park2023background} experimented with sixty-two participants to examine if scope highlighting influences novice programmers' gaze behavior, task performance in speed and correctness, comparing Java code in BlueJ and Java-alike code in Stride. They found that although scope highlighting affects programmers' low-level gaze behaviour, this difference did not persist into high-level comprehension. Villalobos et al.~\cite{villalobos2024experiment} experimented with 33 students with varying programming expertise in Eclipse and found that reviewing C code from different perspectives did not make a difference in performance. 

Most of these studies (except ~\cite{cornelissen2010experiment}) used single-file code snippets and introduced no assistance. The study conducted by Cornelissen et al.~\cite{cornelissen2010experiment} adopted an industry-scale program and assigned 90 minutes for the experiment to elevate its realism; however, it could have overwhelmed some participants and did not log any physiological data for triangulation. 

\subsection{Eye Movement Modeling Examples}

Learning science theories suggest that novices can acquire effective strategies by observing expert behaviors.
Social learning theory~\cite{bandura1977social} describes how modeled behavior can be learned through observation when learners attend to and encode relevant features of the model's performance.
Complementing this view, cognitive apprenticeship~\cite{collins2015cognitive} argues that many critical components of expert performance are tacit and must be made visible so that novices can observe, practice, and gradually internalize them via scaffolding and fading.
In program comprehension, a major portion of this tacit expertise concerns how experts allocate attention and navigate information in complex artifacts such as unfamiliar large codebases.

A closely related line of work in eye-tracking-based instruction is the eye movement modeling examples~\cite{bednarik2018eye}, where learners observe task demonstrations together with a visual overlay of the model's eye movements to guide attention.
This foundational work shows that gaze-based guidance can meaningfully shape attention allocation, but also that the instructional design of the gaze overlay matters and can even hinder learning when it introduces additional cognitive load~\cite{van2009attention}.
Subsequent work demonstrates that specific visualizations, such as spotlight-style guidance, can improve performance and lead to benefits that persist in new and unguided cases~\cite{jarodzka2012conveying}.
This suggests a pathway from attention guidance to transferable perceptual cognitive strategies, and these results motivate careful design choices when using expert gaze as instructional scaffolding.

In the realm of computing education, early evidence shows that the eye movement modeling examples can support source code comprehension.
Bednarik et al~\cite{bednarik2018eye} reported a classroom study where students' comprehension strategies were cued using an intervention that included an expert programmer's eye-movement visualization, yielding improved outcomes relative to baselines.
Emhardt et al~\cite{emhardt2020introducing} further argue that the eye movement modeling examples may support programming learning by establishing joint attention between the model and the learner.
They also highlighted open design questions such as whether models should behave naturally or didactically and how gaze should be visualized.
Building on these prior works, our work investigates whether visualizing expert-derived attention cues inside an IDE helps novices adopt expert-like code reading and navigation strategies in an unfamiliar codebase, and whether such strategies show immediate transfer to a second, comparable codebase without support.

\section{GazePrinter}
\label{sec:gaze-printer}

%\todo{Emma: revisit framing of GazePrinter in relation to the RQs and the focus on visualization of expert gaze}
%
% \todo{Peng: add more details about the development of GazePrinter, e.g., there were more design proposals - done by Peng}

We developed a tool, we call GazePrinter, as an example of how a tool can visualize expert gaze to novices to assist with program comprehension of a new codebase. The design choice was made based on our previous studies, multiple rounds of prototyping, and related work. 
GazePrinter was developed as a plugin in the JetBrains IntelliJ IDE. We chose this platform because there have not been so many SE eye-tracking studies on it. Additionally, when we were planning our study, a third-party plugin CodeGrits~\cite{tang2024codegrits} was released to support eye-tracking data collection on this platform. This further motivated us to make this choice. 

\begin{figure}[htbp]
    \centering
    \includegraphics[width=\linewidth]{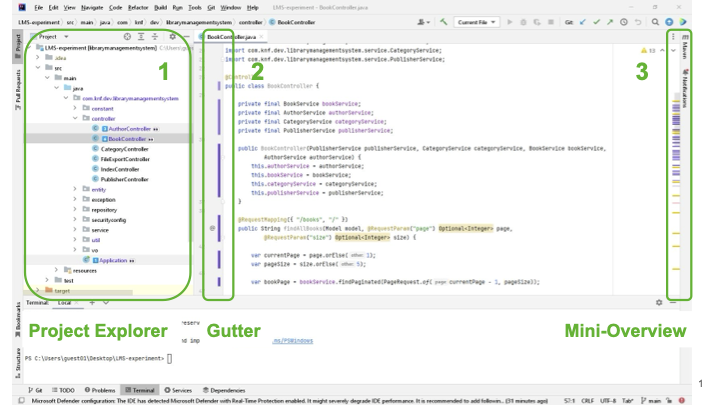}
    \caption{The GazePrinter prototype.}
    \label{fig:gp-prototype}
\end{figure}

\subsection{Visualization Strategy}
Visualizing gaze in a real-world IDE must consider multiple factors. First, the Application Programming Interfaces (APIs) provided by the IDE may limit the areas and levels of customization that we can achieve. Second, the IDE per se is a rich and complex environment, which means that its communication channels are usually already occupied or utilized by its own features and/or assistance provided by other tools. In other words, we may have to share communication channels with other mechanisms or tools. This has both benefits and drawbacks. The benefit is that these channels may have proved to be effective and that programmers may already be accustomed to them. The drawback is that sometimes it may lead to a competition for programmers' attention. Third, for our study, we also want to incorporate some novelty for the research purpose, which in part depends on the strategy that we eventually choose to visualize the gaze. 

Our visualization strategy is the product of the insights gained from our previous work~\cite{kuang2024designing, kuang2024devsurvey} and references to relevant studies in empirical SE~\cite{bednarik2018emme, ahrens2019eyetracking, cheng2022collaborative, saranpaa2023gander} and EMME~\cite{emhardt2020introducing,tunga2023EMMEs} research, taking into account the factors mentioned above.  We explored a variety of design options in specific ways of visualization before arriving at the final design choice. The way in which we can visually link the gaze with a specific code of interest took us multiple rounds of discussion and prototyping. The trials of ideas for this part include coloring the line numbers of the code, attaching an eye icon next to the line of code, and highlighting the code itself (this has already been tested with some users in our previous study~\cite{kuang2024designing}). In the end, we decided to leverage the project explorer, the gutter (instead of the editor, which is somehow crowded and can lead to visual clutter), and the mini-overview areas/features of the IDE, as shown in Figure~\ref{fig:gp-prototype}.

\subsection{Architecture}
\begin{figure}[htbp]
    \centering
    \includegraphics[width=0.85\linewidth]{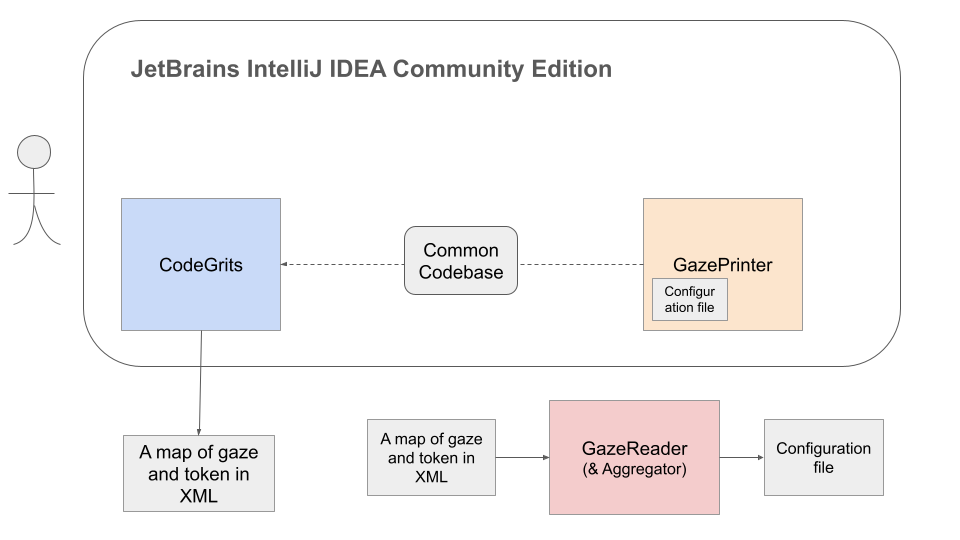}
    \caption{The architecture of GazePrinter.}
    \label{fig:gp-achitec}
\end{figure}

We illustrate GazePrinter's architecture in Figure~\ref{fig:gp-achitec}. We used CodeGRITS to collect model gaze data with five experienced Java programmers on the \gls{LMS} project. However, the gaze data of two experts were excluded because they were incomplete or empty. After that, we processed and aggregated the remaining data to produce a configuration file. We then embedded this file into the GazePrinter plugin and visualized experts' gaze-derived attention as visual assistance in the same codebase, namely, the \gls{LMS} project. 

\subsection{Design and Implementation}

\begin{figure}[htbp]
    \centering
    \includegraphics[width=0.85\linewidth]{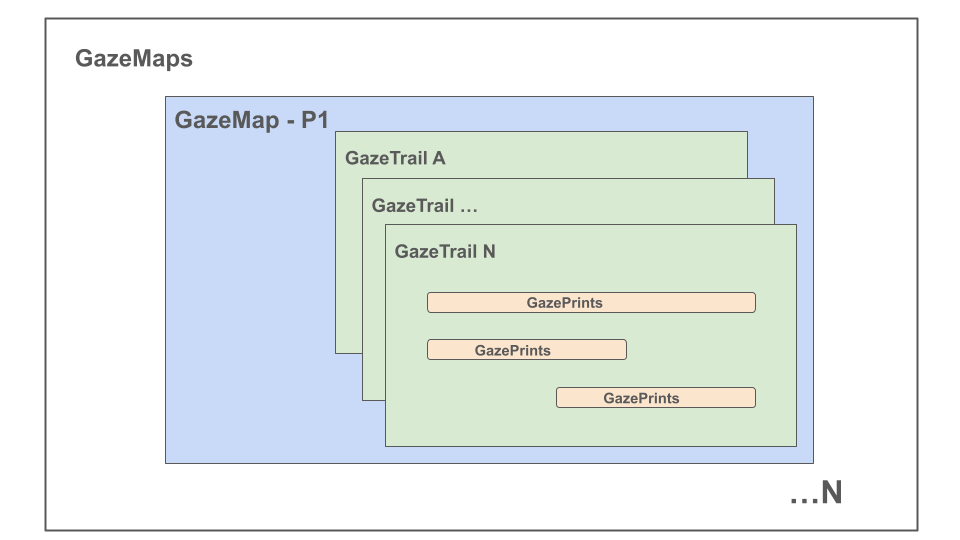}
    \caption{The data structure of GazePrinter.}
    \label{fig:gp-data-struc}
\end{figure}

 We design the core data structure of the GazePrinter tool as in Figure~\ref{fig:gp-data-struc}. We conceptualize programmers' visual attention to code tokens as their gazeprints. Clusters of gazeprints constitute programmers' gaze trails on a certain code file. A group of gaze trails for a codebase becomes a gaze map. By capturing as many programmers' gazes as possible, we build a book of gaze maps that can serve as a guide for newcomers to this code space. 

Our gaze-assistance design is integrated and visualized into three areas of the JetBrains IntelliJ IDE as shown in Figure~\ref{fig:gp-prototype}. Area 1 is the project explorer, in which the top 10 files that received the most attention from expert programmers were highlighted and ranked. Area 2 is the gutter next to the code in the editor. There, we visualize the attention given to the code blocks as vertical color bars. The intensity of these color bars reflects the level of attention that the neighboring code fragments receive. Area 3 is the mini-overview of the selected code file. Within it, we have colored horizontal bars corresponding to the lines of code. The intensity of the color again indicates the degree of attention that has been given to the specific line.

% \vspace{-1cm}
%
\section{Method}
\label{sec:method}

We conducted a mixed-method study to address the research questions listed in Section~\ref{sec:intro}. 
We address \RQ{1} by considering the effect on using GazePrinter with regard to program comprehension efficiency (correctness) and effectiveness (time), and cognitive load.
To address \RQ{2} we consider the effect of using GazePrinter with regard to visual attention measured by comparing file and module reading order, attention distribution, and line-level attention.
For \RQ{3}, we consider the effects of using GazePrinter with regard to the participants confidence in software engineering tasks and their perception of understanding, helpfulness, and learning.
Finally \RQ{4}, we consider the user experience of using GazePrinter.

%We break down \RQ{1} into the following supporting research questions in the context of software engineering (SE) and visualization of expert gaze to novices:
%\begin{itemize}
%    \item \RQ{1.1}: Does it lead to a higher program comprehension efficiency?
%    \item \RQ{1.2}: Does it lead to a higher program comprehension effectiveness?
%    \item \RQ{1.3}: Does it lead to a reduced cognitive load?
%    \item \RQ{1.4}: Does it lead to a different reading behavior?
%\end{itemize}
%We further break down \RQ{2} into the following supporting research questions in the same context:
%\begin{itemize}
%    \item \RQ{2.1}: Does it lead to a higher confidence in SE tasks?
%    \item \RQ{2.2}: Does it lead to a higher perception of understanding?
%    \item \RQ{2.3}: Does it lead to a higher perception of helpfulness?
%\end{itemize}

\begin{figure}[H]
\begin{center}
\includegraphics[width=\linewidth]{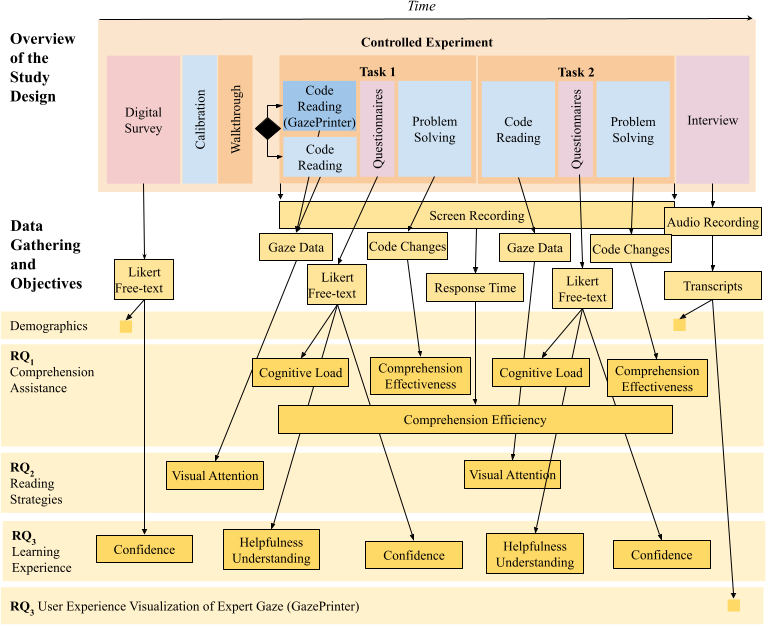};
\end{center}
\caption{Overview of the study design, together with the form and purpose of the data extraction.}
\label{fig:study-design}

\end{figure}

The study design, illustrated in the top of Figure~\ref{fig:study-design}, includes (1) an initial survey aiming to gather details about the previous experience of the participant and confidence in SE tasks (see Section~\ref{sec:survey-design}), (2) a controlled experiment aiming to measure the effect of visualization of expert gaze via GazePrinter (see Section~\ref{sec:experiment-design}), and (3) an interview to gather further participant details along with their perspectives on using GazePrinter and the experiment.
Before the study began, participants were informed about the purpose of the study and informed consent was obtained. At the end of the study, participants were given details about the tasks in the experiment (e.g., codebases used), alongside a compensation of 25 Swiss Francs for their time.  

A replication package for the study is available here: \href{https://doi.org/10.5281/zenodo.18850563}{https://doi.org/10.5281/zenodo.18850563}

\subsection{Survey}
\label{sec:survey-design}

The survey included $22$ questions grouped into seven sections designed for screening participants' eye conditions, understanding details of their educational background, their programming training and experiences, their self-perceived programming proficiency and familiarity with the technical concepts and tool stack selected for the experiment, their self-perceived skill levels in software engineering, and their confidence in software engineering tasks. The survey was implemented as an online form with a combination of structured responses and short free-text responses.

\subsection{Experiment}
\label{sec:experiment-design}

\begin{table}[th]
\caption{Overview of the experiment design.}
\label{tab:expr-design}
\centering
\begin{tabular}{lllll}
\hline
\textbf{Group} & \textbf{Task 1} & \textbf{Task 2} & \textbf{Primary Analysis Focus}\\
\hline
\textbf{Control} & \makecell[l]{Project 1, \\no treatment} &  \makecell[l]{Project 2, \\no treatment} & \makecell[l]{measure natural transfer of behavior\\ delta}\\
\hline
\textbf{Experiment} & \makecell[l]{Project 1,\\\textbf{with treatment}} & \makecell[l]{Project 2,\\no treatment} & \makecell[l]{measure behavior delta \\influenced by the\\ treatment}\\
\hline
 & \makecell[l]{Between-subjects \\comparison of \\initial performance} & \makecell[l]{Within-subjects \\comparison of\\change over time} & \\
 \hline
\end{tabular}
% \vspace{0.2em}
\end{table}

%  \makecell[l]{Compare the deltas to\\ assess if the treatment\\modifies the learning rate}
\begin{table}[ht]
\caption{Summary of the independent and dependent variables of the experiment.}
\label{tab:expr-vars}
\centering
\begin{tabular}{lll}
\hline
%\textbf{RQs} & 
\textbf{Independent Variable} & \textbf{Dependent Variables} & \textbf{Measures} \\
\hline
%\RQ{1.1} & 
\multirow{7}{*}{Visualization of Expert Gaze}  
     & Comprehension Efficiency & Response time \\ 
%\RQ{1.2} &
& Comprehension Effectiveness & Task performance \\
%\RQ{1.3} &
& Cognitive Load & Self-assessment (NASA TLX) \\
%\RQ{1.4} &
& Visual Attention & Gaze distribution \\
%\RQ{2.1} &
& Confidence & Self-assessment (Likert) \\
%\RQ{2.2} &
&  Perceived Understanding & Self-assessment (Likert) \\
%\RQ{2.3} &
&  Perceived Helpfulness & Self-assessment (Likert) \\ 
\hline
\end{tabular}
\end{table}

% \begin{table}[htbp]
% \caption{Independent and Dependent Variables of the Experiment.}
% \label{tab:expr-vars}
% \centering
% \begin{tabular}{lll}
% \hline
% Independent Variables & Dependent Variables & Measurements \\
% \hline
% Gaze-based visual assistance & Speed & Response time \\
% \hline
%  & Understanding & Task performance \\
% \hline
%  & Cognitive Load & NASA TLX score \\
% \hline
%  & Transfer learning & Task performance \\
% \hline
%  & Visual attention & Gaze distribution \\
%  \hline

% \end{tabular}
% % \vspace{0.2em}
% \end{table}

We designed a mixed-factorial experiment, illustrated in Table~\ref{tab:expr-design}, where participants are assigned to a control group or an experiment group in rotation.  
The participants in both groups perform two program comprehension tasks in the same order. The order is deliberately predefined to measure immediate transfer of reading strategies between tasks (\RQ{2}). Because of the design with two groups and two tasks, we can perform both between-subjects and within-subjects analyzes. 
We list the independent and dependent variables in Table~\ref{tab:expr-vars} and provide more details below.

\textbf{Independent Variable: Visualization of Expert Gaze} Participants are provided with one of two alternatives: an IDE with support from GazePrinter (the treatment), or an IDE with no support from GazePrinter. Expert gaze is collected from three researchers (originally five, two excluded in the later phase) with substantial industrial experience working as software engineers and knowledge of the primary programming language Java, and the technical stack. 

% \todo{details about the experts used to configure GazePrinter? - [Peng]Already covered with a paragraph in experimental apparatus. Maybe the question becomes where is the best place to mention this information? - done by Peng}

\textbf{Dependent Variable: Comprehension Efficiency}
We measured efficiency via task response time, recorded by the first author using the built-in timer on a mobile phone during the experiment, and verified it through the screen recording before data analysis. 
The non-parametric Mann-Whitney U test, which disregards whether the response times are normally distributed, is performed to check whether the differences in the response times between the two groups are significant. Accordingly, Cliff's delta is applied to check the effect size of their differences. The uncertainty of the effect size is reported using the confidence interval.

\textbf{Dependent Variable: Comprehension Effectiveness}
We measured effectiveness by assessing task performance. The completion and correctness of the task were manually examined by the first author; a half or full point was assigned to each of the solutions presented by the participant. For instance, for a task of changing a button to a designed color, the participant receives a full point if the button is changed to the right color, but gets a half point if the button is changed to a different color. The between-subject differences of the resulting task scores are again analyzed through the Mann-Whitney U test, Cliff's delta, and confidence interval.

\textbf{Dependent Variable: Cognitive Load}
We measured cognitive load using the NASA TLX form~\cite{hart1988nasa-tlx} after each reading task. Participants were asked to rate six contributing factors of their workload on a scale of 20 and then pick one from each of the fifteen pairs of contributors on the paper questionnaires. The contributing factors included physical demand, mental demand, temporal demand, performance, effort, and frustration. This was completed immediately after participants finished their reading of the codebases. The frequency with which a contributor had been picked became its weight. For each participant, the cognitive load of each task was calculated as the sum of the products of each factor's rating and its weight, divided by 15. We applied the Mann-Whitney U test, Cliff's delta, and confidence interval for the between-subject analysis. 

\textbf{Dependent Variable: Visual Attention}
We measured visual attention through a selection of metrics derived from the gaze distribution of each group. 
Gaze data is extracted from the XML data produced by a third-party plugin CodeGRITS~\footnote{https://github.com/codegrits/CodeGRITS}. From this data, the gaze distribution metrics for each group are computed as follows; (1) we take the maximum hit on any single token of a line of code for each participant, (2) we calculate the mean hit of that particular line across all participants for each group, and (3) we categorize the line hits into five attention levels according to their value ranges and distribution characteristics (e.g., skewness). Based on this gaze distribution metric, we analyze the file reading order, module reading order, attention distribution, and line-level attention of each group, as follows:
\begin{itemize}

\item \textbf{File reading order}: We compose the files viewed by each group as a sequence of file name strings. We calculate the distance between the sequences of each group and the reference sequence, that is, the recommended expert sequence visualized in the codebase via GazePrinter. We use the Dynamic Time Warping algorithm~\cite{berndt1994dtw}, which accounts for speed and length variations of compared sequences, for the similarity calculation. 

\item \textbf{Module reading order}: We map the files in the selected projects to modules guided by the annotations in the Spring Boot framework, which is used by both projects. For files without any annotation, we coded them with the names of their parent folders or the modules of the well-known Model-View-Controller architecture, but with minor variations, e.g., a file may be mapped to `Entity' instead of precisely `Model'. The first and second authors mapped all the files separately and discussed cases where they were unsure. Full consensus was reached between the authors after discussion. We aggregated the file sequences for each group and task into module sequences, using one character to represent each module. We compare the sequences using the normalized Needleman-Wunsch algorithm~\cite{wiki2026needlemanwunsch}. 

\item \textbf{Attention distribution}:
We count the maximum number of times a token is viewed on a line as LineHits for that line. We normalize LineHits by the participant's session duration before being stored, so that participants who took a longer time reading the code do not disproportionately influence the later data aggregation. We take the means of LineHits for each line of each file across all the participants in each group. We then categorize the lines into five grades of attention according to their means of LineHits, using a self-developed algorithm that accounts for the skewness of the distribution of LineHits across participants in that group. 

\item \textbf{Line-level attention}:
We count the number of lines each group looked at in each file, and we compare the line overlap between each pair of groups utilizing the Jaccard overlap index~\cite{wiki2025javvard}. 

\end{itemize}

For confidence, we perform an within-subjects analysis on pre- and post-task confidence level ratings reported by the participants. For perceived understanding and helpfulness, we perform a between-subjects analysis on the degrees of understanding and helpfulness reported by participants. 

\textbf{Dependent Variable: Confidence} 
% \todo{add details about how this data is analyzed - done by Peng}
We categorize typical software development activities into four types: code review, debugging, refactoring, and implementation of new features. This strategy aims to capture the nuanced differences in participants' confidence with respect to different software engineering activities. We gather participants' corresponding confidence in each of these activities before the experiment (in the survey) and after each of the two tasks (on paper questionnaires). The confidence is reported on a scale of 10, with 10 indicating the highest level of confidence. We applied the Mann-Whitney U test for the between-subjects analysis and the Wilcoxon signed-rank test~\cite{wiki2026wilcoxonsigned} for the within-subjects analysis.

\textbf{Dependent Variable: Perceived Understanding}
% \todo{add details about how this data is analyzed - done by Peng}
We ask participants to rate their level of understanding of the codebase after they complete each task. Contexualized in software maintenance, they also report how readable the code is. Both measures are gathered on a scale of 10, with 10 as the maximum positivity. The self-reported understanding reveals the percentage of knowledge acquired about the codebase, as perceived by participants themselves. The readability helps infer whether it notably confounds the perceived understanding. We applied the Mann-Whitney U test for the between-subject analysis. To avoid reporting false positives of significance caused by the multiple comparison problem~\cite{wiki2026multicomparisonsprob}, we applied the Bonferroni correction~\cite{wiki2026bonferronicorrection} to both analyzes. 

\textbf{Dependent Variable: Perceived Helpfulness} 
% \todo{add details about how this data is gathered and analyzed - done by Peng}
Perception of helpfulness is gathered as part of the questionnaire at the end of the experiment, prior to the interview. We applied the Mann-Whitney U test for the between-subjects analysis.

% TODO: incorporate the below text 
%For the experiment, the experiment group had an additional 3-minute walkthrough of the treatment, the GazePrinter tool, after calibration. The entire session lasted about an hour and 15 minutes on average, including logistics. 
%
%The experiment session comprised two tasks, each task containing two parts. The first part was reading a codebase, and the next part was making changes to the codebase to solve the two provided problems. 
%
%The experiment includes initial calibration and walkthrough, for those participants assigned to the treatment (GazePrinter), followed by code reading and code editing tasks interleaved with questionnaires gathering details about cognitive load. 
%
%We employed a variety of tools to collect different formats of data with the purpose of triangulating findings for the study, as illustrated in Figure~\ref{met:experiment-procedure}. 
%
%In between the code reading and problem-solving sessions, we asked participants to rate the sources of mental effort and compare the dimensional contributors to their cognitive load through paper questionnaires. 
%
%We also enabled the screen recorder when participants started their tasks; however, this part of the data was only intended for verification, in case we wanted to better understand a certain part of the other data.

\subsubsection{Selection of Projects}
\label{met:material}

% \todo{explain why there is a need for two projects and not one - done by Peng}

To measure the transfer learning between tasks, we agreed on the necessity of having two different yet comparable projects. Additionally, a second project can act as the baseline. This helps reinforce the findings: a) it can reduce the risk of the results' overfitting to any unique characteristics of the first project, and b) thereby can increase the reliability of our interpretations of the results. 

We selected the following two projects as the study materials: a \gls{LMS} and an \gls{E-comm}. They are open-source projects on GitHub, both with mainly Java code and implemented with the Java Spring Boot framework~\footnote{\url{https://spring.io/projects/spring-boot}} supporting applications with a Model-View-Controller (MVC) architecture. The projects also contain microservices~\footnote{\url{https://microservices.io/}} in the backend and some Bootstrap\footnote{\url{https://getbootstrap.com/}} code in the frontend, and both use the Maven build system\footnote{\url{https://maven.apache.org/}}. The \gls{E-comm} project further uses the H2 database\footnote{\url{https://www.h2database.com/html/main.html}} to store its data. The first author lightly modified both projects so that they could run locally and better suit the experiment. For instance, we updated the Spring framework dependency version in the Maven configuration file in order to build and run the~\gls{LMS}. The source code for these two projects are available on GitHub\footnote{Task 1 - LMS: \url{https://github.com/PengKuang/LMS-experiment}; Task 2 - E-commerce: \url{https://github.com/PengKuang/E-commerce-project}}. 

\begin{table}[ht]
\caption{Overview of properties of the two selected projects.}
\label{tab:codebase-props}
\centering
\begin{tabular}{llrrrr}
\hline
\textbf{Project} & \textbf{Language} & \textbf{Files} & \textbf{Blank} & \textbf{Comment} & \textbf{Line of Code} \\
\hline
LMS (Task 1)                        & Java                            & 39            &438              &0           &1,279 \\
 & HTML                             & 16             &93              &6            &842 \\
 & Maven                            & 1              &4              &0             &85 \\
 & Markdown                         & 1             &22              &0             &20 \\
 & Properties                       & 1              &0              &0              &2 \\
 \hline
& \textbf{SUM:}                    & \textbf{58}            & \textbf{557}              & \textbf{6}           & \textbf{2,228} \\
\hline
E-commerce (Task 2)  & JSP                             & 16            & 218              & 8           & 1,334 \\
& Java                            & 21            & 305             & 64           & 1,116 \\
& Bourne Shell                     & 1             & 34             & 62            & 220 \\
& DOS Batch                        & 1             & 35              & 0            & 153 \\
& XML                              & 2              & 0              & 0            & 102 \\
& Maven                            & 1             & 16              & 0             & 74 \\
& Markdown                         & 1             & 23              & 0             & 72 \\
& SQL                              & 1             & 12              & 7             & 46 \\
& Properties                       & 1              & 7             & 11             & 21 \\
& YAML                             & 1              & 4              & 6             & 12 \\
 \hline
& \textbf{SUM:}                    & \textbf{46}             & \textbf{654}            & \textbf{158}           & \textbf{3,150} \\
\hline
\end{tabular}
% \vspace{0.2em}
\end{table}

We chose these two projects because they are types of applications with which our target participants would have opportunities to interact in their daily lives. Therefore, we expected that the participants should already have some prior knowledge about the functionalities of such systems (a.k.a. application domain). In light of this, we believe that any potential difficulty in understanding the business logic of these applications should be minimal. 
%In other words, the demand for mental effort was mainly associated with the code itself but not with the functionalities of the system. 

\textbf{Codebase Properties.}~We analyze the attributes of the two projects to gather metrics that help us better understand their complexity. 
In our view, the complexity of the code is rooted in both the size and the quality (attributes such as indentation, naming, comments, nesting statements, etc.). 
Therefore, we measure the lines of code (and comments), the number and types of files, and cyclomatic complexity that each project contains.

\begin{table}[htbp]
\centering
\caption{Cyclomatic Complexity of the Two Codebases.}
\label{tab:codebase-complex}
\resizebox{0.75\textwidth}{!}{%
\begin{tabular}{llrrr}
\hline
\textbf{Project} & \textbf{Count} & \textbf{ClassName} (Java files only) & \textbf{Total Complexity} & \textbf{Highest Complexity} \\
\hline
Task 1 - LMS & 0&              Application.java           & 2                  & 1\\
& 1&                     Item.java          & 4                  & 1\\
& 2&         \textbf{AuthorController.java}          & \textbf{11}                  & 2\\
& 3&           \textbf{BookController.java}          & \textbf{12}                  & 2\\
& 4&       CategoryController.java          & 10                  & 2\\
& 5&     FileExportController.java          & 2                  & 1\\
& 6&          IndexController.java          & 1                  & 1\\
& 7&      PublisherController.java          & 10                  & 2\\
& 8&                   Author.java          & 10                  & 1\\
& 9&                     \textbf{Book.java}          & \textbf{24}                  & 1\\
& 10&                Category.java          & 8                  & 1\\
& 11&               Publisher.java          & 8                  & 1\\
& 12&                    Role.java          & 6                  & 1\\
& 13&                    \textbf{User.java}          & \textbf{14}                  & 1\\
& 14&       NotFoundException.java          & 1                  & 1\\
& 15&   SecurityConfiguration.java          & 4                  & 1\\
& 16&       AuthorServiceImpl.java          & 8                  & 2\\
& 17&         BookServiceImpl.java          & 10                  & 2\\
& 18&     CategoryServiceImpl.java          & 6                  & 1\\
& 19&         FileServiceImpl.java          & 7                  & 5\\
& 20&    PublisherServiceImpl.java          & 6                  & 1\\
& 21&         UserServiceImpl.java          & 5                  & 3\\
& 22&                  Mapper.java          & 4                  & 1\\ \hline
&  &                                 & SUM: 173                 & MAX: 5\\
\hline
Task 2 - E-commerce & 0 &       HibernateConfiguration.java            & 3                  & 1\\
& 1&   JtSpringProjectApplication.java             & 1                  & 1\\
& 2&        SecurityConfiguration.java             & 3                  & 1\\
& 3&    AdminConfigurationAdapter.java             & 1                  & 1\\
& 4&     UserConfigurationAdapter.java             & 1                  & 1\\
& 5&               \textbf{AdminController.java}  & \textbf{24}                  & 3\\
& 6&              ErrorController.java             & 1                  & 1\\
& 7&               \textbf{UserController.java}    & \textbf{15}                  & 2\\
& 8&                     cartDao.java              & 5                  & 1\\
& 9&               cartProductDao.java             & 6                  & 1\\
& 10&                 categoryDao.java             & 7                  & 2\\
& 11&                  productDao.java             & 7                  & 2\\
& 12&                     userDao.java             & 9                  & 3\\
& 13&                        Cart.java             & 5                  & 1\\
& 14&                 CartProduct.java             & 8                  & 1\\
& 15&                    Category.java             & 4                  & 1\\
& 16&                     \textbf{Product.java}    & \textbf{16}                  & 1\\
& 17&                        \textbf{User.java}    & \textbf{12}                  & 1\\
& 18&                 cartService.java             & 4                  & 1\\
& 19&             categoryService.java             & 5                  & 1\\
& 20&              productService.java             & 5                  & 1\\
& 21&                 userService.java             & 7                  & 3\\ \hline
&  &                                      & SUM: 149                 & MAX: 3\\
\hline
\end{tabular}
}
\end{table}

According to Table~\ref{tab:codebase-props}, the E-commerce project used in Task 2 has more lines of code (3150 LoC) and more types of languages (10 types). Although its total number of files is less than the Library Management System (LMS) project, it has a greater variety of file types. Also, because the E-commerce project has fewer files but more lines of code, it means the length of its files is longer than those of LMS on average. 
It is reasonable to say that Task 2 is more challenging than Task 1 due to the scale of the code involved. 
However, it is noteworthy that the E-commerce project also contains more comments, which usually enhance the readability and understandability of the code, thereby reducing its complexity to some extent.

Table~\ref{tab:codebase-complex} shows that the two projects share the same number of files with a cyclomatic complexity above 10 (three files of medium complexity between 11 and 20 and one of high complexity at 24), while the total cyclomatic complexity of the LMS project is slightly higher than the E-commerce project. On average, the complexity of the Java files for the LMS project is 7.5, and the complexity of the E-commerce project is 6.8, both falling into the category of low complexity.

Overall, these metrics suggest that these two projects are of comparable complexity. The LMS project exhibits slightly higher cyclomatic complexity but contains less code, whereas the E-commerce project has more code but slightly lower cyclomatic complexity. In other words, the LMS project seems slightly more challenging but involves less reading, while the E-commerce project may be easier but requires more reading. 

% \vspace{2mm}
% \setlength{\fboxsep}{7pt}%box to content distance
% \setlength{\fboxrule}{2pt}%thickness of box
% \fcolorbox{gray!60}{gray!20}{%
%     \parbox{0.88\columnwidth}{%
%         \textbf{Codebase Summary}: Overall, the two codebases share comparable complexity. The code for Task 1 is slightly more challenging. The code for Task 2 is easier to read but requires more reading. 
%         }
% }
% \vspace{2mm}

\subsubsection{Description of Tasks}
\label{met:task}

We designed two tasks (illustrated in Figure~\ref{fig:study-design}), with one code reading and problem solving part each, as follows:
\begin{itemize}
    \item \textbf{Task 1}
    \begin{itemize}
        \item \textbf{Code Reading}: Read the codebase in the \gls{LMS} project with the aim of comprehending the program.
        \item \textbf{Problem Solving}: After guiding the participants on how to run the application from the command line interface. They were given the following two problems to solve:
        \begin{itemize}
        \item Change the color of the ``Add Authors'' button from dark to light.
        \item Make the author table display only the first author instead of all three authors.
        \end{itemize}
        The participants were offered the flexibility to choose to solve either problem first for both tasks. Participants could use Google, ChatGPT, and any other internet resources for this task. It was an `open exam'. 
    \end{itemize}
    \item \textbf{Task 2}
    \begin{itemize}
        \item \textbf{Code Reading}: Read the codebase in the \gls{E-comm} project with the aim of comprehending the program.
        \item \textbf{Problem Solving}: After guiding the participants on how to run the application from the command line interface. They were given the following two problems to solve:
        \begin{itemize}
        \item Change the color of the ``Add Category" button from blue to green.
        \item Sort the category table by the category name in ascending alphabetical order
        \end{itemize}
        The participants were offered the flexibility to choose to solve either problem first for both tasks. Participants were not allowed to use any internet resources while solving these problems. It was a `closed exam'. 
    \end{itemize}
\end{itemize}

\subsubsection{Experimental Apparatus}

\begin{figure}[htbp]
    \centering
    \includegraphics[width=0.75\linewidth]{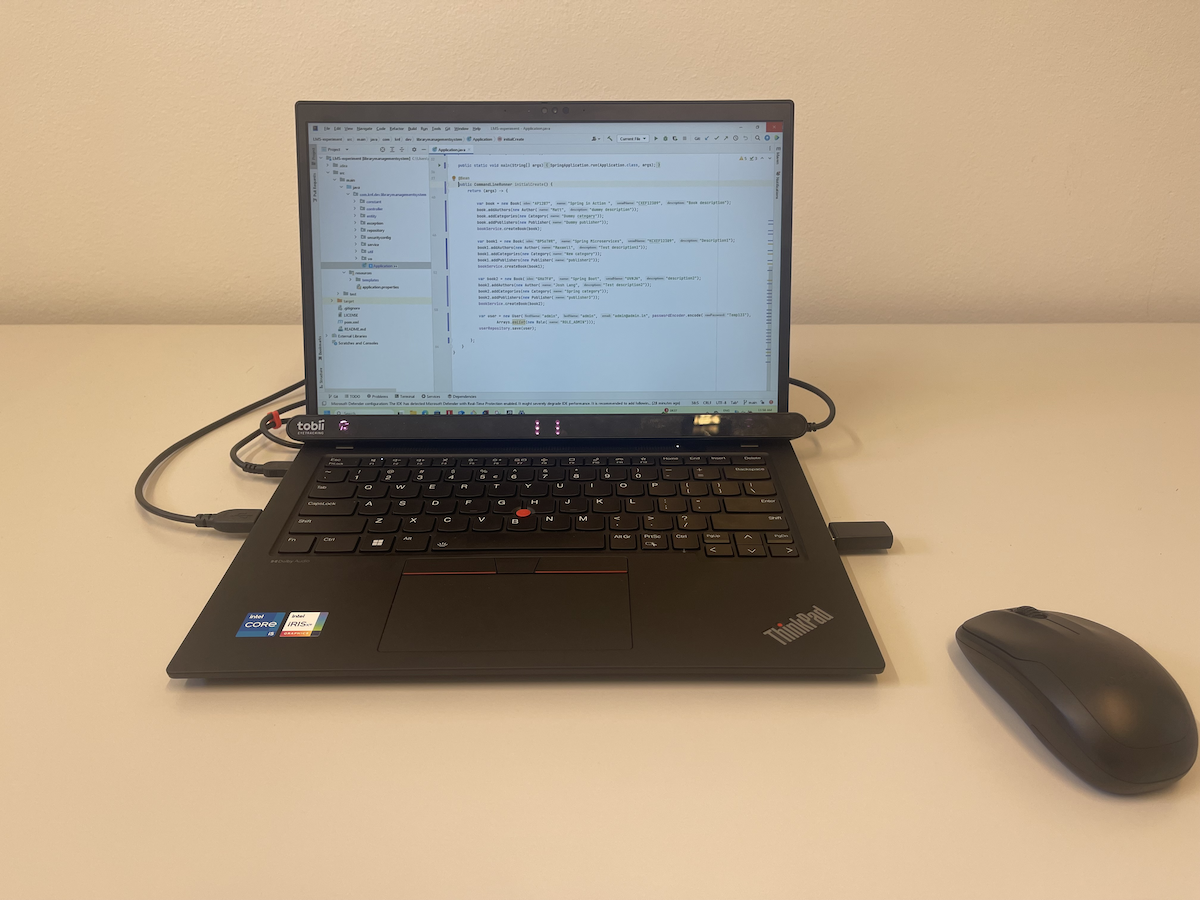}
    \caption{The experimental setup.}
    \label{fig:study-apparatus}
\end{figure}

The experimental setup consists of a Tobii 4C eye tracker and a Lenovo ThinkPad T14s model (as shown in Figure~\ref{fig:study-apparatus}). The Tobii eye tracker was set at 90 Hz, which is sufficient to examine reading behavior~\cite{angele2025low}. The laptop screen size was 14" (317.5 x 226.9 mm), with a resolution of 1680 x 1050 pixels and a refresh rate of 60 Hz. The code was displayed in the light mode of JetBrains IntelliJ Integrated Development Environment (IDE) Community Edition, version 2024.2.5. We used the default JetBrains Mono font at 13 points with a line height set to 1.2 (which translates into 15.6 points for the vertical line spacing).

To configure GazePrinter for the project used in the experiment, we conducted a study with five expert Java programmers at Lund University in Sweden before the experiment. The five experts are researchers specialized in Java compilers and/or Software Development in the Department of Computer Science. They all have substantial experience working as a Software Engineer in the industry. We collected their gaze data as they carried out the same code reading task as the participants for the first project, and used the aggregated gaze from the group to configure how GazePrinter provides assistance to the experiment group.

\subsubsection{Gaze Calibration}
\label{met:calibration}
% \todo[author=Peng]{add accuracy, precision values - done}

Calibration is a necessary step for eye tracking studies to collect high-quality gaze data. Calibration was performed using the Python script provided by Titta~\cite{niehorster2020titta}, an open-source library for eye-tracking studies. We recorded the accuracy and precision, as determined using the standard nine-point validation procedure of Titta, for both eyes of each participant. For the control group, the mean accuracies were 0.12 (left eye) and 0.13 (right eye), and the mean precisions were 0.43 (left eye) and 0.48 (right eye). For the experiment group, the mean accuracies were 0.10 (left eye) and 0.10 (right eye), and the mean precisions were 0.43 (left eye) and 0.42 (right eye). These metrics were all within an acceptable range, and the means of the two groups were comparable.

% TODO: Incorporate
%The calibration for most participants usually took one or two attempts. Occasionally, participants who wore glasses or appeared tense underwent more than two calibration attempts before the results were accepted. We accepted the calibration results when the accuracy and precision were under 1.0 or could not be further improved. 
%
%The participants carried out the tasks on a laptop we provided. We used an eye tracker along with the third-party CodeGRITS plugin~\cite{tang2024codegrits} to collect gaze data from participants during their code-reading sessions. 

\subsection{Interview}
\label{sec:interview-design}

The questions centered on participants' strategies for reading and understanding new codebases. Participants in the experiment group received additional questions about the usefulness and usability of GazePrinter. In the end, the participants were given the opportunity to ask questions about the study. 

%TODO: Incorporate
%The questions of the interview seek to better understand participants' program comprehension strategies, gauge the effect of potential confounding factors, and elicit their perceptions on the expected outcomes. If they belong to the experiment group, they are also asked to share their opinions about the GazePrinter tool, primarily around its usefulness and usability. 

The interviews were recorded using the Voice Memos software built into the first author's work laptop - an Apple MacBook Pro with M1 chip.
The transcripts of the interview recordings were generated using Microsoft Word's Transcribe feature. We sampled 10 percent (N=4) of the transcripts for pilot coding, two from each group, respectively. The sample was selected from the middle part of the data because we deemed the head and tail parts more likely to contain outliers.
The probing questions were still developing during the first interviews, and some of the later interviews tended to be more verbose.

The first and second authors open-coded the pilot transcripts independently and then met to discuss their coding and derive a coding scheme. After that, the two authors adjusted their coding for those four transcripts by applying the coding scheme. Then, they met again to check the coding of both and resolve disagreements on two of the four selected transcripts. The first author continued to adjust the coding for the remaining two of the pilot transcripts with the second author's review. Upon a consensus on the coding between the two authors, the first author coded the remainder of the transcripts. 
First, whenever there was a discrepancy in the codes between the first and second authors, they explained their rationale and evidence to each other. When there was any ambiguity around a participant's answer, they consulted the audio recording to clarify it. Second, while coding the remainder of the transcripts, the first author listened to the corresponding recording at least once for each interview. Third, the last author reviewed the coding results after the first author finished all the coding. The authors discussed the codes and modified them for better clarity where applicable.

We map the codes into a hierarchy consisting of theme, topic, category, and code. We count the frequency of occurrence for the codes. We also report the ratio of their counts to the total counts, converted into a percentage. For mutually exclusive codes under a category, the sum of their percentages equals 100\%. Otherwise, they are non-mutually exclusive codes. If the sum is larger than 100\%, it indicates that multiple codes exist within a single participant's answer. If it is less, it suggests that some participants did not report anything relevant under that category. As the semi-structured interview was placed at the end of the experiment session, we did not get to ask every participant all the probing questions of our interest due to time constraints. We had to be selective, apart from asking the primary set of questions. Therefore, we selectively combined the candidate questions that were most interesting to us at the time, mainly around their regular strategy for reading and understanding a new codebase for the first time, and some peripheral questions on this aspect. In addition to that, for participants from the experiment group, we delved more into questions concerning the tool GazePrinter, such as its usefulness, usability, and potential improvements. This results in there being codes named ``Not asked'' under certain categories. 

We approach participants’ general process of reading and understanding new codebases from the dimensions of strategy, tools and resources, time, and experience. For strategy, we synthesize the categories from the literature with some new patterns reported in our study. For tools and resources, we categorize the results from the participants into AI, conventional tools, and human resources. For time, we group them into different periods, including Year, Month, Week, Day, Hour, and Minute. We then
report how many times participants mentioned such periods relative to the project scale (if compared) presented in the experiment. For experience, we asked participants to share their experience from the perspective of sentiment and perceived causes of difficulties.
\section{Results}
\label{sec:results}

We conducted our study at the Federal Institute of Technology Zurich (ETH Zurich) in Switzerland during the first quarter of 2025. We recruited 40 participants based in Zurich via the assistance of the DecSiL Lab at ETH (N=32) using the University Registration Center for Study Participants\footnote{\url{https://www.uast.uzh.ch/index}}), and other channels such as posters, emails, LinkedIn, Slack, and word of mouth (N=8). The experiment lasted around 60 minutes per session.

\subsection{Participant Details}

Overall, the participants in this experiment are Java beginners (Section~\ref{sec:participant-proficiency}) and new to the technical concepts and frameworks used in the projects examined (Section~\ref{sec:participants-familiarity}). However, they have a background in computing-related fields (Section~\ref{sec:participant-education}) and programming experience in other languages. The distribution of participants with these characteristics is balanced between the two groups. 

%Below, we present the survey results centered on participants' demographics.

\subsubsection{Educational Background.}
\label{sec:participant-education}
The majority (29/38; control: bachelor N=6, master N=8; experiment: bachelor N=5, master N=10) of participants were at the time of the study pursuing a master's or bachelor's degree in Computer Science (N=11, control N=6, experiment N=5) and Software Engineering (N=3), or a neighboring discipline that involves a computing component. It is noteworthy that although they need to write code for their coursework, their coding is mainly for data analysis. 

\subsubsection{Programming Experience and Java Proficiency.}
\label{sec:participant-proficiency}
On average, the control group has 4.0 years of general programming experience and 0.7 years of paid programming experience, while the experiment group has 4.7 and 0.6 (as shown in Table~\ref{tab:bkg-prog-exp}). Both groups have a relatively large standard deviation due to the presence of one outlier in each of the groups. Despite that, the two groups appear to have comparable programming experience, both in general and paid. In terms of Java proficiency, the two groups share the same level at 2.2 on a Likert scale of 5, with a minor difference in standard deviation. When asked to compare their programming proficiency with a programmer with ten years of experience, the majority of the participants rated themselves as worse or much worse.

\begin{table}[h]
\caption{Overview of programming experience per experiment group.}
\label{tab:bkg-prog-exp}
\centering
\begin{tabular}{lcccc}
\hline
\textbf{Group} & \makecell{\textbf{General Programming} (total) \\(mean years, std.)} & \makecell{\textbf{Paid Programming}\\(mean years, std.)}  & \makecell{\textbf{Java Proficiency}\\(mean level, std.)}  & \makecell{\textbf{Compared to a}\\\textbf{10-year-experience}\\\textbf{programmer}\\(mean level, std.)} \\
\hline
Control                & 4.0 (\textpm 3.4)           &0.7 (\textpm 1.9)             & 2.2 (\textpm 0.8)   & 1.6 (\textpm 0.7)\\
Experiment             & 4.7 (\textpm 4.4)           &0.6 (\textpm 1.2)             & 2.2 (\textpm 0.4)   & 1.3 (\textpm 0.7)\\
\hline

\end{tabular}
% \vspace{0.2em}
\end{table}

\subsubsection{Familiarity with the Technical Concepts and Stack.}
\label{sec:participants-familiarity}
Most of the participants (on average, 75\% of each group report a familiarity below 2.0 on a Likert scale of 5.0, control N=19, experiment N=19; a value of 2.0 corresponds to the answer "Not familiar") respond that they are not familiar with or have never used the technical concepts and stack used in the projects for this experiment. This includes the model-view-controller architecture (control 1.9, experiment 2.0), microservices (control 1.8, experiment 1.9), the Java Spring framework (control 1.7, experiment 1.7), and the Maven building tool (control 1.5, experiment 1.5). The participants (21/38) also predominantly use Visual Studio Code for programming and have little or no experience with the JetBrains IntelliJ IDE that is used in the experiment.

\begin{table}[h]
\caption{Summary of proficiency (max=10) in programming activity per group.}
\label{tab:bkg-prog-act}
\centering
\begin{tabular}{lcccc}
\hline
\textbf{Group} & \makecell{\textbf{Code Review} \\(mean, std.)} & \makecell{\textbf{Debugging}\\(mean, std.)}  & \makecell{\textbf{Refactoring}\\(mean, std.)}  & \makecell{\textbf{Implement new features}\\(mean, std.)} \\ 
\hline
Control                & 4.7 (\textpm 2.3)      &4.6 (\textpm 2.8)      & 3.5 (\textpm 2.3)  & 4.7 (\textpm 2.5)\\
Experiment             & 5.0 (\textpm 2.3)      &4.8 (\textpm 2.1)      & 4.0 (\textpm 2.1)  & 5.2 (\textpm 2.2)\\
\hline

\end{tabular}
% \vspace{0.2em}
\end{table}

\subsubsection{Frequency Working with New Codebases.}
\label{sec:participants-codebases}
Among the participants (N=36) who gave valid responses, 89\% report that they worked with a maximum of 5 codebases in the past six months. On average, the control group worked on 2.8 codebases while the experiment group worked on 3 codebases. Surprisingly, from those (N=28) who can recall the size of the codebases they worked on, 57\% of the codebases contain 1000 or more lines of code. It can reach tens of thousands across hundreds of files. Most of these codebases (17/23) are closed-source. 

%\vspace{2mm}
%\setlength{\fboxsep}{7pt}%box to content distance
%\setlength{\fboxrule}{2pt}%thickness of box
%\fcolorbox{gray!60}{gray!20}{%
%    \parbox{0.88\columnwidth}{%
%        \textbf{Result summary 1}: Overall, the participants of this experiment are Java beginners and new to the technical concepts and frameworks used in the projects being examined. However, they have a background in computing-related fields and programming experience in other languages. The distribution of participants with these characteristics is balanced between the two groups. 
%        }
%}
%\vspace{2mm}

% \input{figs/coding-results-pc-booktaps}

\subsubsection{Codebase Comprehension - Strategy}

The most shared strategies are top-down (57.5\%), integrated (32.5\%), and bottom-up (25\%). Among the instances of the top-down strategy, the most common one is depth-first comprehension, accounting for 42.5\%. It refers to the pattern that the participant starts with a preferred entry point and then follows the relationship chain of the code of interest, such as the method calls and variable declarations, to establish a deep understanding of the codebase. Another variation of this strategy is architecture-following, demonstrated by 7.5\% of the participants. It means the participants read the code in a manner that aligns with the codebase's logical structure or underlying framework, e.g., Model-View-Controller. Interestingly, 7.5\% of the participants also mentioned that a walk-through by a person who is more experienced with the codebase would be a great opener. 

In comparison, 25\% of the participants described their comprehension strategies that fall into the bottom-up category. These participants stated that they typically scan all the files and then read them in a natural order. Within the file, they also tend to read the code line by line. A small percentage, at 12.5\% of participants, indicated that they read and understand the code according to the demand of the task at hand; they usually try not to read and understand the entire codebase at once, but to the extent that allows them to do the task. While this is voluntary, another participant claimed that he read code incrementally because he could not direct his attention for too long, and thus had to plan his reading based on iterative, short sessions. Action-assisted, tool-assisted, and no strategy are the new categories we observed in this study. 22.5\% of the participants shared that they rely on executing the code (e.g., running the test cases), making some changes to the code (e.g., building a toy feature), and/or clicking around as a user to understand new codebases. 

% \clearpage
\setlength{\tabcolsep}{2pt}
\begin{longtable}{c c c c c l}
\caption{The coding results for codebase comprehension}
\label{table:interview-pc}\\

% \begin{tabular}{c c c c c l}
\toprule
\textbf{Theme} & \textbf{Topic} & \textbf{Category} & \textbf{Code} & \textbf{Count} & \textbf{Percentage} \\
\midrule
\endfirsthead

\toprule
\textbf{Theme} & \textbf{Topic} & \textbf{Category} & \textbf{Code} & \textbf{Count} & \textbf{Percentage} \\
\midrule
\endhead

\multirow[c]{15}{*}{Comprehension} 
& \multirow{8}{*}{Strategy} 
& Top-down & Depth-first & 17 & \deltacell{42.5} \\
& & & Architecture-following & 3 & \deltacell{7.5} \\
& & & People-assisted & 3 & \deltacell{7.5} \\

& & Bottom-up & Breadth-first & 10 & \deltacell{25} \\

& & As-needed & Task-driven & 3 & \deltacell{7.5} \\
& & & Incremental & 2 & \deltacell{5} \\

& & Action-assisted & Execution & 1 & \deltacell{2.5} \\
& & & Making changes & 5 & \deltacell{12.5} \\
& & & User perspective & 3 & \deltacell{7.5} \\

& & Tool-assisted & AI & 3 & \deltacell{7.5} \\
& & & debugger, search engine, etc. & 3 & \deltacell{7.5} \\

& & Integrated & Miscellaneous & 13 & \deltacell{32.5} \\

& & No strategy & Acquaintance hopping & 5 & \deltacell{12.5} \\

\cmidrule(lr){2-6}
% \endfoot

& \multirow{7}{*}{Tools \& Resources} 
& AI & ChatGPT & 19 & \deltacell{47.5} \\
& & & Co-pilot, Gemini or Claude & 4 & \deltacell{10} \\

& & Conventional & Google & 14 & \deltacell{35} \\
& & & StackOverFlow & 3 & \deltacell{7.5} \\

% \midrule
% \endfoot

& & Human & Friends & 4 & \deltacell{10} \\
& & & Colleagues & 3 & \deltacell{7.5} \\
& & & Supervisors & 2 & \deltacell{5} \\
& & & Unspecified & 6 & \deltacell{15} \\

& & / & Not asked & 6 & \deltacell{15} \\

\cmidrule(lr){2-6}
% \endfoot

& \multirow[t]{15}{*}{Time} 
& Year & Unspecified & 1 & \deltacell{2.5} \\

\endfoot

& & \makecell[c]{Month \\\footnotesize (1-6 months)} & Larger size & 2 & \deltacell{5} \\

% \midrule
% \endfoot

& & \makecell[c]{Week \\\footnotesize (1-3 weeks)} & Larger size & 2 & \deltacell{5} \\
& & & Smaller size & 1 & \deltacell{2.5} \\
& & & Unspecified & 1 & \deltacell{2.5} \\

& & \makecell[c]{Day \\\footnotesize (0.5-6 days)} & Larger size & 1 & \deltacell{2.5} \\
& & & Unspecified & 7 & \deltacell{17.5} \\

& & \makecell[c]{Hour \\\footnotesize (0.5-11 hours)} & Larger size & 2 & \deltacell{5} \\
& & & Comparable size & 3 & \deltacell{7.5} \\
& & & Smaller size & 2 & \deltacell{5} \\
& & & Unspecified & 7 & \deltacell{17.5} \\

& & \makecell[c]{Minute \\\footnotesize (1-29 minutes)} & Comparable size & 1 & \deltacell{2.5} \\
& & & Smaller size & 3 & \deltacell{7.5} \\
& & & Unspecified & 4 & \deltacell{10} \\

& & / & Not asked & 10 & \deltacell{25} \\

\cmidrule(lr){2-6}
% \endfoot

& \multirow{9}{*}{Experience} 
& Sentiment & Mixed (temporal) & 6 & \deltacell{15} \\
& & & Mixed (code attributes) & 3 & \deltacell{7.5} \\
& & & Mixed (availability of support) & 3 & \deltacell{7.5} \\
& & & Positive & 5 & \deltacell{12.5} \\
& & & Neutral/Manageable & 10 & \deltacell{25} \\
& & & Neutral (temporal) & 2 & \deltacell{5} \\
& & & Neutral (availability of support) & 1 & \deltacell{2.5} \\
& & & Negative & 5 & \deltacell{12.5} \\
& & & Not asked & 9 & \deltacell{22.5} \\

& & Source of Difficulty & Code size & 8 & \deltacell{20} \\
& & & Code quality & 4 & \deltacell{10} \\
& & & Code complexity & 3 & \deltacell{7.5} \\
& & & Identifying critical or relevant code & 3 & \deltacell{7.5} \\
& & & Building a mental representation & 3 & \deltacell{7.5} \\
& & & Unfamiliarity & 4 & \deltacell{10} \\
& & & First-time effect & 8 & \deltacell{20} \\
& & & Not asked & 13 & \deltacell{32.5} \\
\bottomrule
\endlastfoot
% \end{tabular}
\end{longtable}

As AI is becoming a dominant type of tool assistance in programming, it is not surprising that 7.5\% of the participants specified that they use AI to help them with that, in addition to the 7.5\% leveraging conventional tools like debuggers and search engines. Again, interestingly, 12.5\% of the participants claimed that they have no strategy. This can perhaps be attributed to their inexperience in programming and working with code at scale. However, since these participants also disclosed that they usually start with or stop at any code fragments that appeared familiar to them (e.g.,  a programming language or concept encountered before), we coined their reading behavior as `Acquaintance hopping`. Finally, a total of 32.5\% of the participants are counted as `Integrated` if they revealed two or more of the above-mentioned strategies, excluding `No strategy`.

%\vspace{2mm}
%\setlength{\fboxsep}{7pt}%box to content distance
%\setlength{\fboxrule}{2pt}%thickness of box
%\fcolorbox{gray!60}{gray!20}{%
%    \parbox{0.88\columnwidth}{%
%        \textbf{Finding 6}: A set of diverse comprehension strategies exists %among the participants. While classic strategies dominated, new strategies emerged with new tools such as AI. 
%        }
%}
%\vspace{2mm}

\subsubsection{Program Comprehension - Tools and Resources}

For participants asked about tools and resources used for program comprehension (N=34), we found that AI had already become the primary tool (57.5\%) that participants resort to for programming problems, making conventional tools less frequently used, with 35\% for Google and 7.5\% for StackOverflow. Although asking humans, such as friends, classmates, workmates, and supervisors, still accounts for 37.5\% in total, participants often did not mention it in their first response to the interviewer's question but instead positioned it as a secondary resource or backup plan, with the assumption of having access to them. 

%\vspace{2mm}
%\setlength{\fboxsep}{7pt}%box to content distance
%\setlength{\fboxrule}{2pt}%thickness of box
%\fcolorbox{gray!60}{gray!20}{%
%    \parbox{0.88\columnwidth}{%
%        \textbf{Finding 7}: The majority of the participants use AI to %support code comprehension, but over a third of participants still rely on humans, at least as backup support.
        % }
%}
%\vspace{2mm}

\subsubsection{Program Comprehension - Time}

Most participants (N=22) indicated that they would need hours (N=14) or days (N=8) to comprehend a new codebase. Their comprehension goal appeared to be to (1) establish an overview of the codebase, or to (2) accomplish a thorough understanding of the codebase so that they would be able to introduce changes in it, for instance, to fix a bug or implement something new. Some mentioned (N=12) that it took them weeks (N=4) or minutes (N=8), which revealed a large discrepancy if not contextualized. Furthermore, two participants talked about having spent months, and one reported the need for one year. Six participants explicitly stated that whether or not participants have prior knowledge with respect to the codebase makes a difference. If it were an entirely new codebase, their time spent largely depended on the size of the codebase that they had to work on. 15\% of the participants specified that they usually worked with codebases of a larger size than the ones presented in the experiment. 12.5\% mentioned they usually worked with codebases of a comparable size, and 15\% also mentioned codebases of a smaller size. 50\% of the participants did not specify the size of the code. Because usually the size of a codebase (total lines of code) was not explicitly indicated either in a code editor or a Git repository, it was not so easy for the participants to give an accurate number or even a rough estimate, especially given that the vast majority of them were novices. %
So in the later phase, we asked participants to compare the program scale they normally worked with against the size of the projects presented in the experiment instead. 

%\vspace{2mm}
%\setlength{\fboxsep}{7pt}%box to content distance
%\setlength{\fboxrule}{2pt}%thickness of box
%\fcolorbox{gray!60}{gray!20}{%
%    \parbox{0.88\columnwidth}{%
%        \textbf{Finding 8}: The time needed for comprehending a new codebase varies among the participants, ranging from minutes to a year. 
%        }
%}
%\vspace{2mm}

% \input{figs/code-reading-modeling}
% no pain no gain, pivot point

\subsubsection{Program Comprehension - Experience}
 
Surprisingly, not many participants reported negative sentiment towards tackling a new codebase. They indicated that because they enjoyed programming, working with new code did not particularly incur negative emotions from within them. Some of these participants further explained that they did not attach a specific feeling to the code or task, positive or negative; it is normally manageable. However, some also reasoned that it was perhaps because the scale or nature of the code they had worked with so far was not so challenging due to limited experience in programming (still a learner). This explains what we observed from listening to the answers from some of the relatively more experienced participants. Surprisingly, the latter cohort mentioned negative sentimental words such as ``overwhelming'' and ``frustrating'' more often. This was perhaps linked to the fact that they had worked on codebases with a more realistic scale and complexity. However, some participants did report both negative and positive sentiments. They described the process as challenging or overwhelming at the beginning, but gradually became less so and even rewarding once they started to understand what they had read. It seems to us that this so-called point of ``understanding it'' is when programmers' feelings pivoted, either from negative to positive or from negative to neutral. Occasionally, participants stated they never felt negatively about it, but the opposite, simply because they had a passion for programming.

%\vspace{2mm}
%\setlength{\fboxsep}{7pt}%box to content distance
%\setlength{\fboxrule}{2pt}%thickness of box
%\fcolorbox{gray!60}{gray!20}{%
%    \parbox{0.88\columnwidth}{%
%        \textbf{Finding 9}: The participants shared mixed sentiment with new codebases. There exists a pivot point for the shift of contrasting feelings among some participants. 
%        }
%}
%\vspace{2mm}

\subsection{Comprehension Assistance: Efficiency, Effectiveness, and Cognitive Load (\RQ{1})}

%Overall, the designed gaze-based tool assistance did not statistically influence participants' response time, task performance, and cognitive load. However, it suggested a weak or uncertain effect on the reduction of response time and cognitive load, which requires larger samples to clarify.

We collect participants' response time recorded by the first author, task performance in points, and their self-reported cognitive load through completion of the NASA TLX questionnaire. 

\subsubsection{Descriptive summary}
On average, the control group spent a longer total time completing the two tasks, but also received slightly higher points than the experiment group; the control group also reported a higher cognitive load. As shown in Table~\ref{tab:task-time-score-cogload}, for Task 1, the control group spent 1115 seconds completing the task, received 0.24 points, and reported a cognitive load of 12.11, while the experiment group spent 1127, received 0.20, and reported 10.78. For task 2, the control group spent 979 seconds, received 0.18 points, and reported a cognitive load of 11.98, while the experiment group spent 859, received 0.16, and reported 10.17.

\begin{table}[h]
\caption{Summary of response time, task performance \& cognitive load per task and group.}
\label{tab:task-time-score-cogload}
\centering

\begin{tabular}{llccc}
\hline
\textbf{Task} & \textbf{Group} & \makecell{\textbf{Response Time} \\(mean sec., std.)} & \makecell{\textbf{Task Performance} \\(mean, std.)}  & \makecell{\textbf{Cognitive Load} \\(mean, std.)} \\
\hline
Task 1 & Control                & 1115 (\textpm 110)           &0.24 (\textpm 0.31)             & 12.11 (\textpm 3.18)         \\
 & Experiment                     & 1127 (\textpm 109)           &0.20 (\textpm 0.27)             & 10.78 (\textpm 2.35)           \\
\hline
Task 2 & Control              & 979 (\textpm 171)          &0.18  (\textpm 0.22)            & 11.98 (\textpm 2.97)        \\
 & Experiment                      & 859 (\textpm 189)           &0.16  (\textpm 0.24)            & 10.17 (\textpm 2.53)         \\
\hline
\end{tabular}

% \vspace{0.2em}
\end{table}

\subsubsection{Statistical Analysis}
Overall, we did not see a significant difference between the two groups among these three measures. In other words, the intervention of our gaze-based tool assistance did not significantly help participants in the experiment group reduce their time completing the tasks, increase their task performance, or mitigate their cognitive load. Below, we provide an explanation of the statistical details for each measure. 

\begin{table}[h]
\caption{Summary of response time per task and group.}
\label{tab:time-stats}
\centering
\resizebox{\linewidth}{!}{
\begin{tabular}{llcccccc}
\hline
 \textbf{Task} & \textbf{Group} & \textbf{Sample} & \makecell{\textbf{Response Time} \\(mean sec.)} & \textbf{Std.} & \makecell{\textbf{Mann-Whitney U} \\(p-value < 0.05)} & \makecell{\textbf{Cliff’s delta} \\(>0.147 small)} & \textbf{Confidence Interval} \\
\hline
T1-Read. & Con. & 19  & 518    & \textpm 106  & \footnotesize U = 205.500,  &  0.139 & \footnotesize mean diff = 28.821,\\
             & Exp. & 19  & 546  & \textpm 84  & \footnotesize p-value = 0.424 &   & \footnotesize 95\% CI = [-28.421, 88.421]  \\
\hline
T1-Solv. & Con. & 19  & 597    & \textpm 14  & \footnotesize U = 170.500,  &  -0.055 & \footnotesize mean diff = -15.809,\\
             & Exp. & 19  & 581  & \textpm 69  & \footnotesize p-value = 0.553 &   & \footnotesize 95\% CI = [-50.526, 6.316]  \\
\hline
T2-Read. & Con.  & 19   & 398  & \textpm 143 & \footnotesize U = 138.500,   & -0.233 & \footnotesize mean diff = -56.974, \\
               & Exp. & 19  & 341  & \textpm 128 & \footnotesize p-value = 0.219 &  & \footnotesize 95\% CI = [-138.947, 28.421]   \\
\hline
T2-Solv. & Con.  & 19   & 581  & \textpm 49 & \footnotesize U = 220.500,   & -0.222 & \footnotesize mean diff = -63.479, \\
               & Exp. & 19  & 518  & \textpm 163 & \footnotesize p-value = 0.136 &  & \footnotesize \textbf{95\% CI = [-145.263, 3.158]} \\
\hline
\end{tabular}
}
% \vspace{0.2em}
\end{table}
As shown in Table~\ref{tab:time-stats}, the Mann-Whitney U tests did not reveal any statistically significant differences in response time between the control and experiment groups across all tasks (all p >0.05). The Cliff's delta for effect size estimation suggests their differences are likely to be negligible, although the experiment group usually took less time to respond to the tasks. The confidence intervals also indicate that the experiment group tends to require less time performing the tasks, as they mostly lean towards the negative end. Because they all contained the zero value, it implies that their differences are not significant. However, for the problem-solving parts of both tasks, there is a potentially small (and meaningful) effect if with a larger sample, as their confidence interval upper bounds are just slightly above zero, especially for Task 2. 

\begin{table}[htbp]
\caption{Summary of task performance (points) per task and group.}
\label{tab:score-stats}
\centering
\resizebox{\linewidth}{!}{
\begin{tabular}{llcccccc}
\hline
\textbf{Task} & \textbf{Group} & \textbf{Sample} & \makecell{\textbf{Task Performance} \\(mean)} & \textbf{Std.} & \makecell{\textbf{Mann-Whitney U} \\(p-value < 0.05)} & \makecell{\textbf{Cliff’s delta} \\(>0.147 small)} & \textbf{Confidence Interval} \\
\hline
T1 & Con. & 19  & 0.24    & 0.31  & \footnotesize U = 169.500,  &  -0.061 & \footnotesize mean diff = -0.039, \\
             & Exp. & 19  & 0.20  & 0.27  & \footnotesize p-value = 0.645 &   & \footnotesize 95\% CI = [-0.224, 0.145]  \\
\hline
T2 & Con.  & 19   & 0.18  & 0.22 & \footnotesize U = 172.000,   & -0.047 & \footnotesize mean diff = -0.013, \\
               & Exp. & 19  & 0.16  & 0.24 & \footnotesize p-value = 0.618 &  & \footnotesize 95\% CI = [-0.158, 0.132]   \\
\hline
% Total (T1 + T2)  & Con.  & 19   & 0.41  & 0.49 & \footnotesize U = 177.000,   & -0.019 & \footnotesize mean diff = -0.052, \\
%                & Exp. & 19  & 0.36  & 0.41 & \footnotesize p-value = 5.489e-01 &  & \footnotesize 95\% CI = [-0.329, 0.224]   \\
% \hline

\end{tabular}
}
% \vspace{0.2em}
\end{table}
Similarly, Table~\ref{tab:score-stats} demonstrates that there is no significant difference between the two groups in terms of their task performance (all p > 0.05). Although the experiment group tends to receive fewer points, evidenced by the negative mean differences and Cliff's deltas, participants of this group do not perform differently from those of the control group to a significant degree. Both groups also received lower scores in Task 2 compared to Task 1. This is perhaps because they could not access the internet in the second task.

\begin{table}[h]
\caption{summary of cognitive load per task and group.}
\label{tab:cogload-stats}
\centering
\resizebox{\linewidth}{!}{
\begin{tabular}{llcccccc}
\hline
\textbf{Task} & \textbf{Group} & \textbf{Sample} & \makecell{\textbf{Cognitive Load} \\(mean)} & \textbf{Std.} & \makecell{\textbf{Mann-Whitney U} \\(p-value < 0.05)} & \makecell{\textbf{Cliff’s delta} \\(>0.147 small)} & \textbf{Confidence Interval} \\
\hline
T1 & Con. & 19  & 12.11    & \textpm 3.18  & \footnotesize U = 8246.000,  &  -0.068 & \footnotesize mean diff = -3.065,\\
             & Exp. & 19  & 10.78  & \textpm 2.35  & \footnotesize p-value = 0.170 &   & \footnotesize \textbf{95\% CI = [-8.538, 2.355]}  \\
\hline
T2 & Con.  & 19   & 11.98  & \textpm 2.97 & \footnotesize U = 8055.000,   & -0.089 & \footnotesize mean diff = -4.154, \\
               & Exp. & 19  & 10.17  & \textpm 2.56 & \footnotesize p-value = 0.104 &  & \footnotesize \textbf{95\% CI = [-9.344, 1.030]}   \\
\hline
\end{tabular}
}
% \vspace{0.2em}
\end{table}
Table~\ref{tab:cogload-stats} seemingly aligns with the hypothesis that our designed tool assistance can help reduce the cognitive load for reading the code. However, we again did not observe a significant difference (all p > 0.05) between the control and experiment groups in this measure, despite that the experiment group reported a lower cognitive load on average for both tasks. This is evidenced by both the negative Cliff's deltas and mean differences. Although the directions of the confidence intervals are consistent with the hypothesis, and their near-zero upper bounds suggest a weak or uncertain effect, it should be interpreted cautiously and needs more data to clarify. 

\vspace{2mm}
\setlength{\fboxsep}{7pt}%box to content distance
\setlength{\fboxrule}{2pt}%thickness of box
\fcolorbox{gray!60}{gray!20}{%
    \parbox{0.88\columnwidth}{%
        \textbf{Finding 1}: Weak indication of a reduction in response time and cognitive load for the experiment group.
        }
}
\vspace{2mm}

\subsection{Reading Strategies (\RQ{2})}

% - coverage of the files and code \\
% - distance or similarity against the recommendation \\
% attention map of an example file
We compare the gaze behaviour of the control and experiment groups with the experts and between each other, considering the file reading order, module-level reading order, attention distribution, and line-level attention.

\begin{table}[htbp]
    \caption{DWT distance distribution per group}
    \label{tab:dwt_dist_stats}
    \centering
    \resizebox{\textwidth}{!}{%
    \begin{tabular}{|l | c | c | c | c | c| c | c | c | c |}
    \hline
    \textbf{Group} & \textbf{Mean} & \textbf{Std} & \textbf{Min} & \textbf{Max} & \textbf{Shapiro-Wilk} & \textbf{Bartlett’s} & \makecell{\textbf{Student's t-test} \\ (p-value < 0.05)} & \textbf{Cohen's d} & \textbf{Confidence Interval}\\
    \hline
    Control & 63.26 & 24.55 & 22.0 & 118.0
    & 0.812 & \multirow{2}{*}{\makecell{stat=0.165,\\ p=0.685}} & \multirow{2}{*}{\makecell{stat=2.351,\\ \textbf{p-value: 0.021}}} & \multirow{2}{*}{\makecell{0.773}} & \multirow{2}{*}{[0.159, 1.465]}\\
    \cline{1-7}
    Experiment & 45.15 & 22.31 & 1.0 & 88.0 & 0.924 &  &  &  &\\
    \hline
    \end{tabular}
    }
    % \vspace{0.2em}
\end{table}

\subsubsection{File Reading Order}
\label{res:dwt-dist}
\begin{figure}[htbp]
    \centering
    \includegraphics[width=0.8\linewidth]{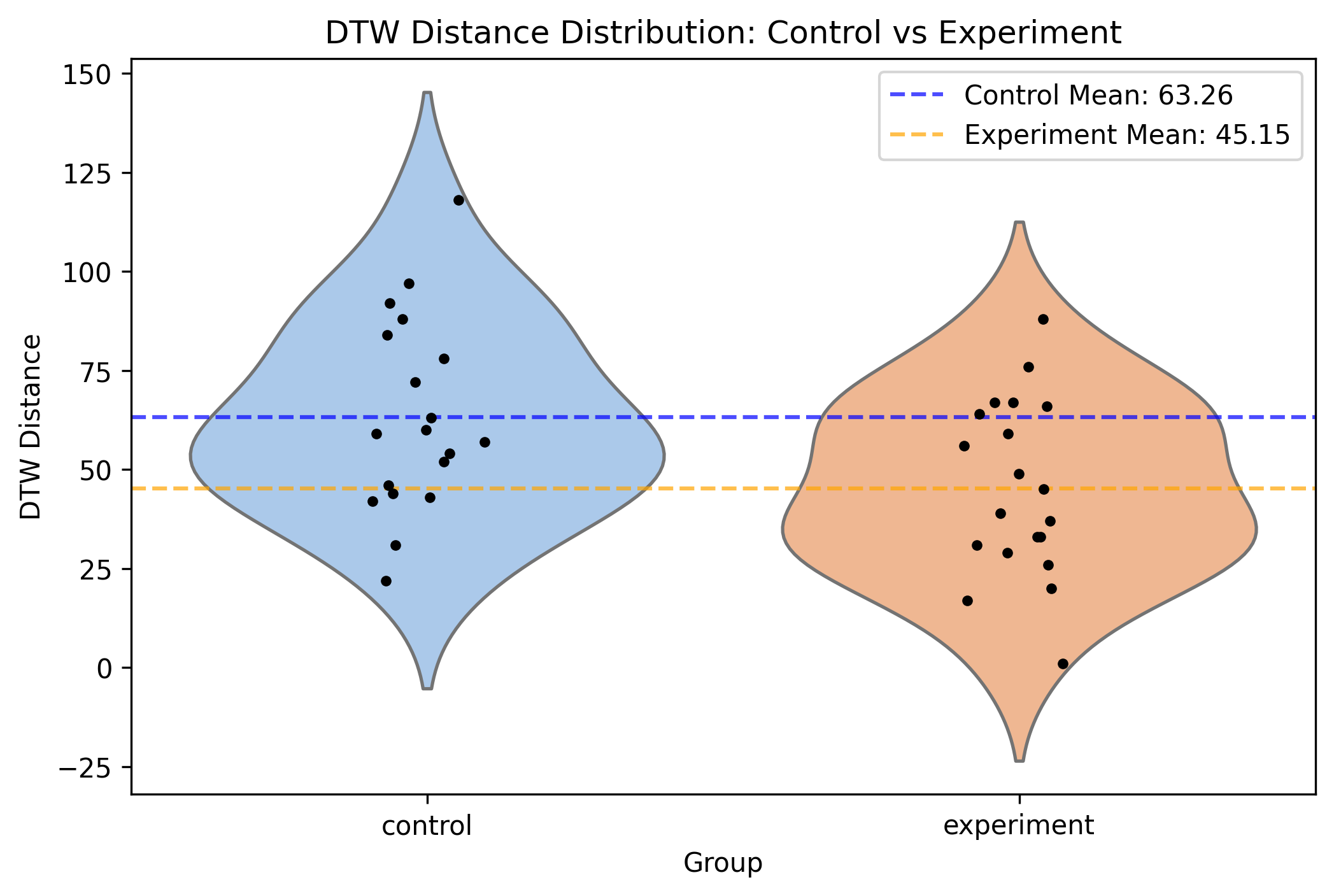}
    \caption{Comparison of the DTW Distance Distribution}
    \label{fig:grp-dwt-dist-violin}
\end{figure}

The descriptive statistics of the control and experiment groups' DTW distances (from the expert group) and their test results are shown in Table~\ref{tab:dwt_dist_stats}. Figure~\ref{fig:grp-dwt-dist-violin} shows the distribution of the data points in these two groups. 
Since the sample size is small, we use the Shapiro-Wilk test to examine its normality. Both groups have a p-value higher than 0.05, implying both are normally distributed. For choosing the appropriate t-test, we examine whether the two groups have roughly equal variances using Bartlett’s test. The differences between their variances are not significant, as the p-value is 0.685. We therefore use Student's t-test to evaluate the statistical significance of the DTW distance differences between the two groups. The p-value of 0.021 indicates that the difference between their means is significant. With a Cohen's d of 0.773 (0.5 as medium and 0.8 as large), we can speculate that the difference has a meaningful effect size. The non-zero confidence interval suggests that this difference has a real, positive effect, although with some uncertainty corresponding to the wide interval. 

\vspace{2mm}
\setlength{\fboxsep}{7pt}%box to content distance
\setlength{\fboxrule}{2pt}%thickness of box
\fcolorbox{gray!60}{gray!20}{%
    \parbox{0.88\columnwidth}{%
        \textbf{Finding 2}: 
        The file reading order of the experiment group is significantly different from that of the control group.
        }
}
\vspace{2mm}

%
% TODO: don't forget about this comment
%
%\todo{what about the comparison to the Expert group? -- because here it is already comparing the two groups' delta to the expert group}

% \todo{Why isn't figure 7 referenced in the text?}
% \todo{include a findings box for the summary}
% \todo{fix the caption of Table 11 which does not include that it is DTW in the table}

\subsubsection{Module Reading Order}

\begin{table}[htbp]
\caption{Similarity and distance of the module sequences between groups.}
\label{tab:similarity}
\centering
\begin{tabular}{lcccc}
\hline
\textbf{Comparison} & \textbf{Group} & \textbf{Task} & \textbf{Similarity} & \textbf{Distance} \\
\hline
Between-subjects & Control vs. Expert & 1 & 0.45 & 0.55 \\
Between-subjects & Experiment vs. Expert & 1 & 0.62 & 0.38 \\
Within-subjects & Control & 1 vs. 2 & 0.34 & 0.66 \\
Within-subjects & Experiment & 1 vs. 2 & 0.50 & 0.50 \\
\hline
\end{tabular}
% \vspace{0.2em}
\end{table}

% gp_ref:  ASCSLCUEEESRCGRCPSGESLCEETPSSRLCSSESSE
% seq1_s1_ctl:  ACCEECEEDCCUESSSPSCEELCSRLXSRLLSRLLPLSSLLPPLGSRTSEPLRGELPPSGLEEP
% seq2_s1_exp:  CCAEUEECSCCSLECSSCSEEEDSSRLSPRXRTLRRSE
% seq1_s2_ctl: CCDERRERRECRLECESASCSSTPPLPP
% seq2_s2_exp: CCRDRRCESRERECSECLAESSTPPPG
% Needleman–Wunsch similarity and distance (seq1_s1_ctl vs seq0_gp): 0.4453125 0.5546875
% Needleman–Wunsch similarity and distance (seq2_s1_exp vs seq0_gp): 0.618421052631579 0.3815789473684211
% Needleman–Wunsch similarity and distance (seq1_s1_ctl vs seq1_s2_ctl): 0.34375 0.65625
% Needleman–Wunsch similarity and distance (seq2_s1_exp vs seq2_s2_exp): 0.5 0.5

According to Figure~\ref{tab:similarity}, for Task 1, the experiment group has a higher similarity of 0.62 and a corresponding lower distance of 0.38 to the expert group, while the control group has a similarity of 0.45 and a distance of 0.55. This suggests that the experiment group exhibits a pattern more similar to the expert group than the control group, potentially due to the introduction of the GazePrinter intervention. Comparing Task 2 with Task 1 among the same groups of participants, the experiment group still has a higher similarity at 0.50 and a lower corresponding distance at 0.50, while the control group has a similarity at 0.34 and a distance at 0.66. This indicates that perhaps the GazePrinter intervention has led to a stronger immediate transfer of reading strategy among participants in the experiment group. They tend to inherit the code-reading pattern from Task 1 to Task 2 at a high level.

\vspace{2mm}
\setlength{\fboxsep}{7pt}%box to content distance
\setlength{\fboxrule}{2pt}%thickness of box
\fcolorbox{gray!60}{gray!20}{%
    \parbox{0.88\columnwidth}{%
        \textbf{Finding 3}: 
        The module reading order of the experiment group is more similar between Task 1 and Task 2 compared to the control group.
        }
}
\vspace{2mm}

\subsubsection{Attention Distribution}

\begin{figure}[h]
    \centering
    \includegraphics[width=0.98\linewidth]{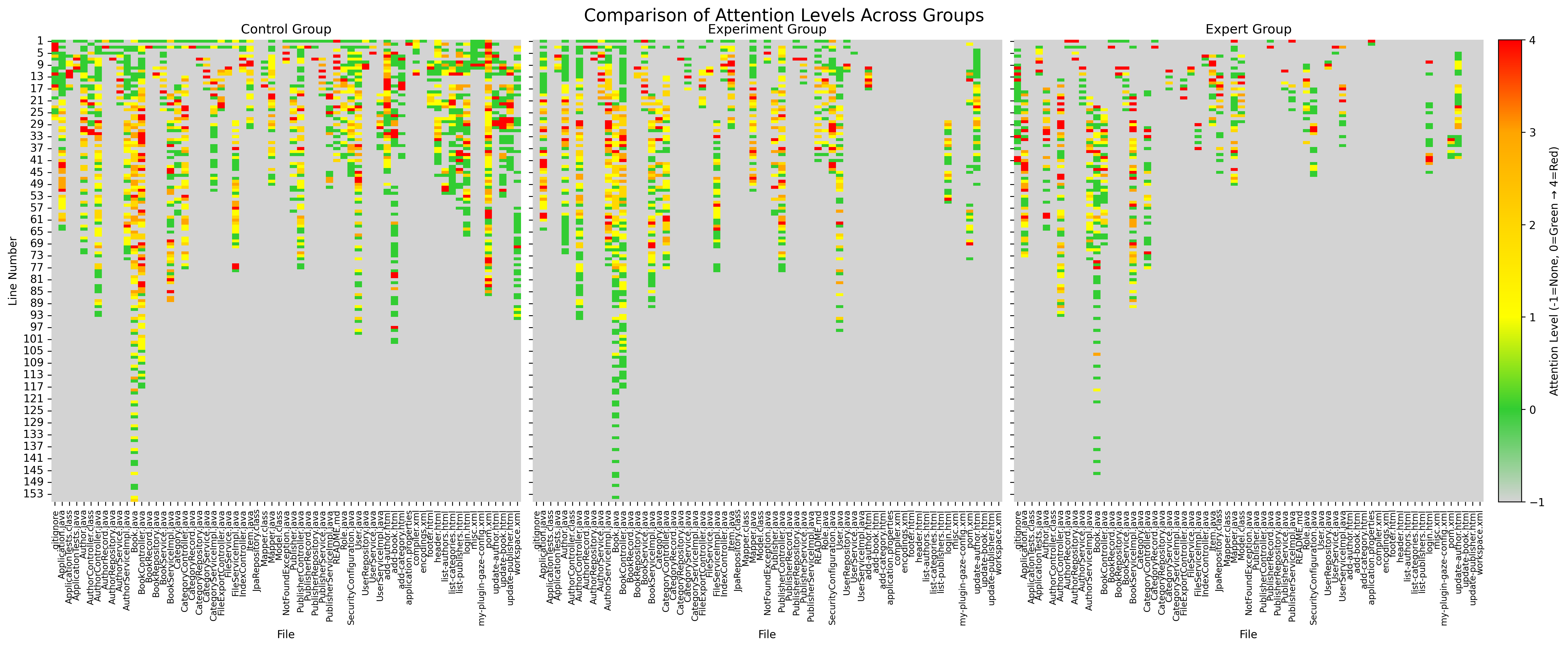}
    \caption{Attention Distribution Per File Per Group}
    \label{fig:attention-all}
\end{figure}

Figure~\ref{fig:attention-all} illustrates the attention distribution of the groups for the files in the study. Participants in the control group covered the most files (N=19, 63 files, and 3.32 files per participant). Participants in the experiment group viewed fewer files (N=19, 38 files, and 1.90 files per participant) compared with the control group. The files they viewed overlap more closely with the expert group, corroborating what we found in the sequence analysis~\ref{res:dwt-dist}. Looking at the top right corners of the subplots, it is reasonable to say that participants in the experiment group seem to have skipped some files that experts did not look at (the grey gap near the top right corner shared by the experiment and expert groups), while those in the control group tend to go through as many files as possible. This difference may have been influenced by the presence of our plugin GazePrinter in the experiment group.

\vspace{2mm}
\setlength{\fboxsep}{7pt}%box to content distance
\setlength{\fboxrule}{2pt}%thickness of box
\fcolorbox{gray!60}{gray!20}{%
    \parbox{0.88\columnwidth}{%
        \textbf{Finding 4}: 
        The attention distribution of the experiment group is closer to the expert group than that of the control group.
        }
}
\vspace{2mm}

\subsubsection{Line-level Attention Comparison}

\begin{table}[h]
\caption{Number of lines looked across files per group.}
\label{tab:line_stats}
\centering
\begin{tabular}{lccc}
\hline
\textbf{Group} & \textbf{Mean Line Count} & \textbf{Min Line Count} & \textbf{Max Line Count} \\
\hline
Control & 24 & 0 & 90 \\
Experiment & 15 & 0 & 91 \\
Expert & 9 & 0 & 58 \\
\hline
\end{tabular}
\end{table}

We count the number of lines each group looked at in each file, shown in Table~\ref{tab:line_stats}. On average, the expert group viewed the fewest lines, while both the experiment and the control groups viewed many more. However, even though both the control and experiment groups share the composition of mostly novices, the average number of lines they viewed differs to a large extent. This disparity may be explained by the fact that the experiment group has been influenced by the intervention of GazePrinter. They focused on a small set of files instead of trying to cover all files. All three groups skipped certain files. In terms of the maximum lines viewed, both the control and the experiment groups have a comparable number (90 vs. 91). However, experts only read 58 lines at most in a file. This is in line with what has been said about them in the literature that they read code more selectively~\cite{aljehane2021codeelement}. 

\begin{figure}[H]
    \centering
    \includegraphics[width=0.7\linewidth]{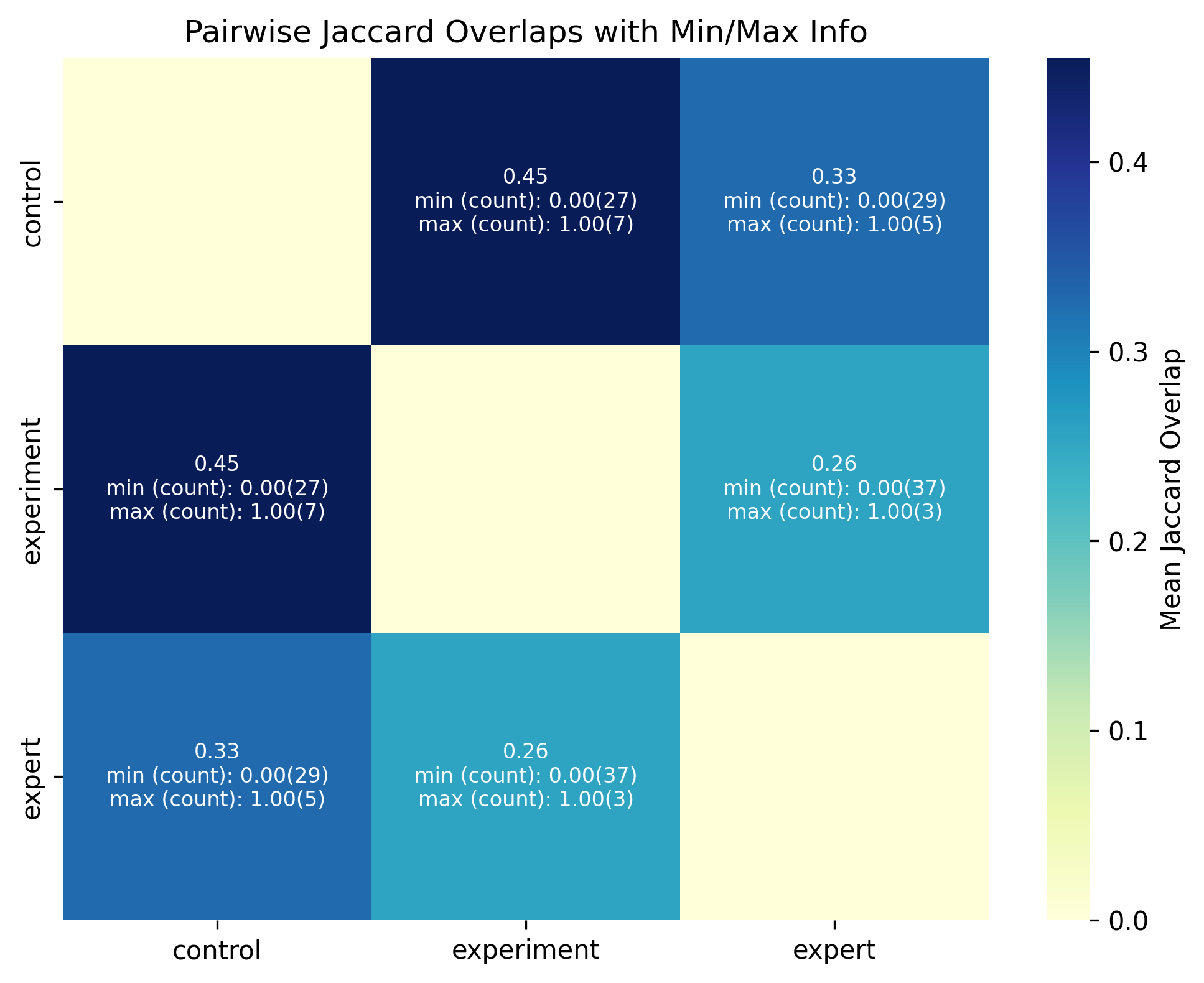}
    \caption{Line-level Overlap between Groups}
    \label{fig:group-line-overlap}
\end{figure}

We further compare the line overlap between each pair of groups, as demonstrated by Figure~\ref{fig:group-line-overlap}. Since the control group covered the vast majority of the files in the project, and the expert group only looked at a very small set of files selectively, these two groups share a high overlap in the matrix with a Jaccard score of 0.33 (sharing zero overlap in 29 files and full overlap in 5 files). Similarly, as the experiment group covered fewer files than the control group, it actually shares a lower score of 0.26 with the expert group (sharing zero overlap in 37 files and full overlap in 3 files). Because both the control and experiment groups consist of novices who tend to read code line by line and in an exhaustive manner, their gaze has a higher chance of overlapping on the line level compared with the expert group. Therefore, these two groups share an overlap score of 0.45 (sharing zero overlap in 27 files and full overlap in 7 files). 

%\vspace{2mm}
%\setlength{\fboxsep}{7pt}%box to content distance
%\setlength{\fboxrule}{2pt}%thickness of box
%\fcolorbox{gray!60}{gray!20}{%
%    \parbox{0.88\columnwidth}{%
%        \textbf{Result summary 3}: Overall, the experiment group read fewer files %compared with the control group, aligning more closely with the expert group. On average, the experts read the fewest lines of code, and the experiment group read fewer lines of code than the control group. 
%        }
%}
%\vspace{2mm}

\vspace{2mm}
\setlength{\fboxsep}{7pt}%box to content distance
\setlength{\fboxrule}{2pt}%thickness of box
\fcolorbox{gray!60}{gray!20}{%
    \parbox{0.88\columnwidth}{%
        \textbf{Finding 5}: 
        On average, the experiment group read fewer lines of code than the control group, and closer to the expert group.
        }
}
\vspace{2mm}

\subsection{Learning Experience (\RQ{3})}

We report participants' learning experiences by combining quantitative and qualitative measures.

\subsubsection{Experiment Results: Confidence, Understanding, and Helpfulness}

% \todo{Peng: add details about the confidence and helpfulness dependent variables here - done by Peng}

We present the results on participants' self-reported confidence, understanding of the codebases, and helpfulness of Task 1 to Task 2.  

\textbf{Confidence.}~According to Table~\ref{tab:learning-confidence-stats}, both groups held comparable levels of confidence across the SE activities before performing the tasks, except in code review, wherein the experiment group showed noticeably higher confidence. Before the tasks, the control group was most confident in debugging, whereas the experiment group was in code review. After Task 1, the control group had the highest confidence in code review, a shift from debugging. For the experiment group, their highest confidence remained in code review, but expanded to debugging as well. After Task 2, both groups reported the highest confidence in code review.  

% \todo{Peng: add statistic for confidence and tasks - done by Peng, added to another table}

\begin{table}[h]
\caption{Summary of confidence (max=10) in programming activity across the experiment.}
\label{tab:learning-confidence-stats}
\centering
\begin{tabular}{llcccc}
\hline
\textbf{Task} & \textbf{Group} & \makecell{\textbf{Code Review} \\(mean, std.)} & \makecell{\textbf{Debugging}\\(mean, std.)} & \makecell{\textbf{Refactoring}\\(mean, std.)}  & \makecell{\textbf{Implement new features}\\(mean, std.)} \\ 
\hline

Before Task1 & Control & 4.89 (\textpm 2.49)  & 5.11 (\textpm 2.92)     & 3.84 (\textpm 2.99)  & 5.0 (\textpm 2.81) \\
& Experiment             & \textbf{5.42} (\textpm 2.59)      & 5.0 (\textpm 2.36)      & 3.95 (\textpm 2.14)  & 5.05 (\textpm 2.88)\\
\hline

After Task1 & Control                & 5.32 (\textpm 2.08)      & 4.58 (\textpm 2.57)     & 4.21 (\textpm 2.02)  & 4.74 (\textpm 2.18)\\
 & Experiment             & \textbf{5.89} (\textpm 2.33)     & \textbf{5.89} (\textpm 2.42)      & 5.26 (\textpm 2.45)  & 5.32 (\textpm 2.54)\\
\hline

After Task2 & Control                & 5.21 (\textpm 2.44)      & 4.63 (\textpm 2.67)     & 4.26 (\textpm 2.35)  & 4.16 (\textpm 2.34)\\
 & Experiment             & \textbf{5.16} (\textpm 2.22)      & 4.79 (\textpm 2.27)      & 3.84 (\textpm 2.01)  & 4.58 (\textpm 2.27)\\
\hline

\end{tabular}
% \vspace{0.2em}
\end{table}

% std to be added
\begin{table}[h]
\caption{Summary of between-subjects analysis per programming activity (* p < 0.003 with Bonferroni correction, 0.05/16).}
\label{tab:learning-confidence-comparison}
\centering
\begin{tabular}{llcccc}
\hline
\textbf{Task} & \textbf{Group} & \makecell{\textbf{Code Review} \\(Btw-sub.)}  & \makecell{\textbf{Debugging}\\(Btw-sub.)} & \makecell{\textbf{Refactoring}\\(Btw-sub.)}   & \makecell{\textbf{Implement new features}\\(Btw-sub.)} \\ 
\hline

Before Task1 & Control & \multirow{2}{*}{p = 0.471}  & \multirow{2}{*}{p = 1.0}    & \multirow{2}{*}{p = 0.626}  & \multirow{2}{*}{p = 0.988}\\
& Experiment             &      &      &  &\\
\hline

After Task1 & Control               & \multirow{2}{*}{p = 0.368}      & \multirow{2}{*}{p = 0.126}     & \multirow{2}{*}{p = 0.237}  & \multirow{2}{*}{p = 0.616}\\
 & Experiment             &     &      &  &\\
\hline

After Task2 & Control             & \multirow{2}{*}{p = 1.0}      & \multirow{2}{*}{p = 0.803}   & \multirow{2}{*}{p = 0.605}  & \multirow{2}{*}{p = 0.617}\\
 & Experiment            &      &     &  &\\
\hline

\end{tabular}
% \vspace{0.2em}
\end{table}

% \todo{Peng: Review the method and results to make sure the correction with the Bonferroni method is covered - done by Peng}

Looking at Table~\ref{tab:learning-confidence-comparison}, we cannot see a statistical difference in confidence between the two groups across all tasks and activities. This is the case both before and after applying the Bonferroni correction~\cite{wiki2026bonferronicorrection} for a potential multiple comparisons problem~\cite{wiki2026multicomparisonsprob} (as we are comparing the same dataset in four different aspects/activities). Nevertheless, their differences appear larger across all SE activities in Task 1, particularly in debugging and refactoring, indicated by the pattern that the smallest p-value of each column (corresponding to each SE activity) all sit in the row of Task 1. Such a pattern suggests that the presence of our tool may have exacerbated the disparity between the two groups in confidence. 

To sum up, for the control group, there was a directional shift in their confidence from debugging at the beginning to code review after completing the tasks. For the experiment group, they consistently reported the highest confidence in code review across all tasks; accompanying this, a temporal expansion of their confidence was developed into debugging after completing Task 1. 

% \begin{figure}[H]
%     \centering
%     \includegraphics[width=0.7\linewidth]{imgs/group_confidence_line_graph.pdf}
%     \caption{Mean Confidence across Tasks per Group.}
%     \label{fig:group-confidence}
% \end{figure}

% Source - https://tex.stackexchange.com/a/397214
% Posted by Mark Wibrow
% Retrieved 2026-03-06, License - CC BY-SA 3.0

% \documentclass[tikz,margin=5]{standalone}

% \begin{document}

\begin{figure}[t]
\centering
    
\begin{tikzpicture}
\begin{groupplot}[
    width=0.45\linewidth,
    group style={
        group size=2 by 2,  % 2 columns x 2 rows
        horizontal sep=1cm,
        vertical sep=1.8cm
    },
    ymin=0, ymax=10, 
    ytick={0,1,...,10},
    xtick={0,1,...,2},
    xlabel=Task,
    ylabel={Confidence [max=10]},
    error bars/.cd, 
    % y dir=both,
    % y explicit,
    error bars/y dir=both,
    error bars/y explicit,
    error bars/error mark=-,
    legend pos=north west,
    legend style={draw=none, font=\small}
]

% ------------------ Plot 1: Code Review ------------------
\nextgroupplot[title={Code Review}]
\addplot+[
  skyblue, mark=*, mark options={skyblue, scale=0.75},
] table [x=x, y=y, y error=error, col sep=comma, row sep=newline] {
x,y,error
0,4.895,1.201
1,5.316,1.004
2,5.211,1.176
};
\addplot+[
  orange, mark=*, mark options={orange, scale=0.75},
] table [x=x, y=y, y error=error, col sep=comma, row sep=newline] {
x,y,error
0,5.421,1.248
1,5.895,1.123
2,5.158,1.069
};
\addlegendentry{control}
\addlegendentry{experiment}

% ------------------ Plot 2: Debugging ------------------
\nextgroupplot[title={Debugging}]
\addplot+[
  skyblue, mark=*, mark options={skyblue, scale=0.75},
] table [x=x, y=y, y error=error, col sep=comma, row sep=newline] {
x,y,error
0,5.105,1.409
1,4.579,1.237
2,4.632,1.287
};
\addplot+[
  orange, mark=*, mark options={orange, scale=0.75},
] table [x=x, y=y, y error=error, col sep=comma, row sep=newline] {
x,y,error
0,5.000,1.136
1,5.895,1.168
2,4.789,1.096
};
% No legend here to avoid duplicates

% ------------------ Plot 3: Refactoring ------------------
\nextgroupplot[title={Refactoring}]
\addplot+[
  skyblue, mark=*, mark options={skyblue, scale=0.75},
] table [x=x, y=y, y error=error, col sep=comma, row sep=newline] {
x,y,error
0,3.842,1.439
1,4.211,0.972
2,4.263,1.179
};
\addplot+[
  orange, mark=*, mark options={orange, scale=0.75},
] table [x=x, y=y, y error=error, col sep=comma, row sep=newline] {
x,y,error
0,3.947,1.035
1,5.263,1.179
2,3.842,0.967
};

% ------------------ Plot 4: Implementing New Features ------------------
\nextgroupplot[title={Implementing New Features}]
\addplot+[
  skyblue, mark=*, mark options={skyblue, scale=0.75},
] table [x=x, y=y, y error=error, col sep=comma, row sep=newline] {
x,y,error
0,5.000,1.354
1,4.737,1.052
2,4.158,1.128
};
\addplot+[
  orange, mark=*, mark options={orange, scale=0.75},
] table [x=x, y=y, y error=error, col sep=comma, row sep=newline] {
x,y,error
0,5.053,1.386
1,5.316,1.224
2,4.579,1.093
};

\end{groupplot}
\end{tikzpicture}

\caption{Mean confidence across tasks per group. Before Task 1 is represented as Task 0.}
\label{fig:group-confidence}

\end{figure}
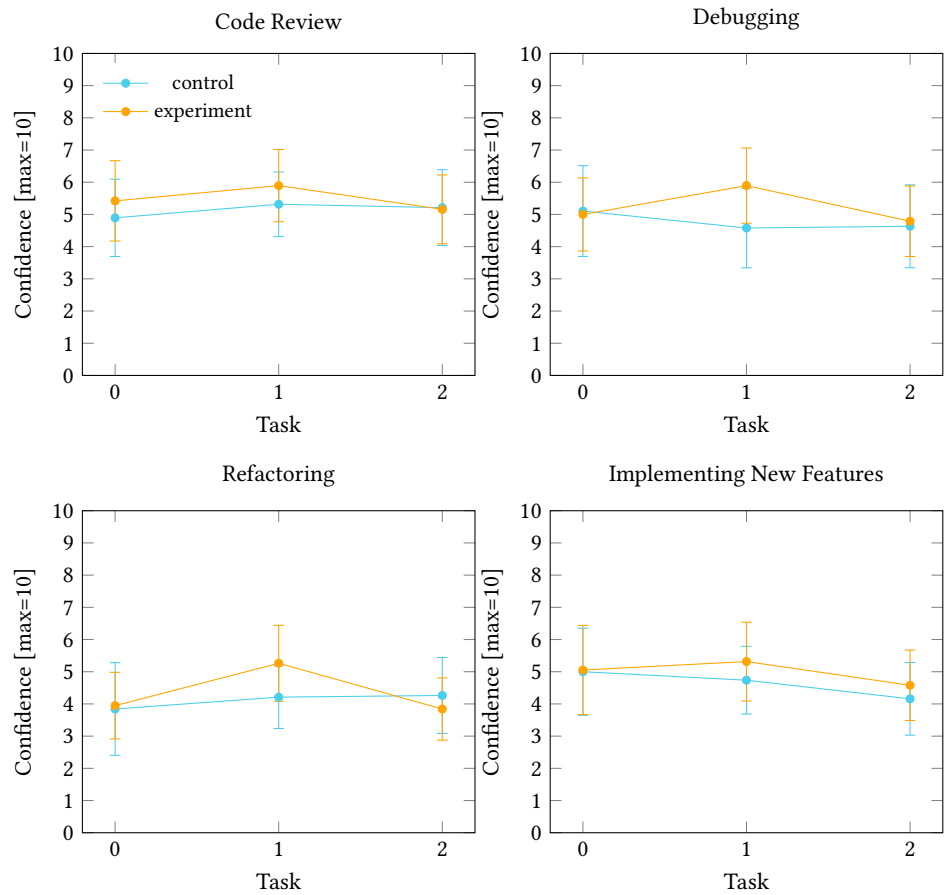
% \end{document}

Next, we delve into both groups' mean confidence changes over tasks in each SE activity. As per Figure~\ref{fig:group-confidence}, the control group slightly grew their confidence in code review and refactoring after completing all the tasks, but decreased confidence in debugging and implementing new features (implementation for short hereafter). Their confidence plunged in debugging after Task 1 and bounced back a bit after Task 2, but continuously dropped in implementation. For the experiment group, their confidence peaked after completing Task 1 and declined after completing Task 2 in all SE activities. This pattern is particularly obvious in debugging and refactoring - the former increased by 17.8\% and the latter increased by 33.2\% after completing Task 1. In spite of that, their confidence in both activities decreased after completing Task 2, to a value slightly lower than at the beginning.

% \todo[author=Peng]{Update stats in the within-subject table and add texts for interpretation - done by Peng}

\begin{table}[h]
\caption{Summary of within-subjects analysis per programming activity (* p < 0.002 with Bonferroni correction, 0.05/24). Before Task 1 is represented as T0, Task 1 as T0, and Task 2 as T2.}
\label{tab:learning-confidence-with-subject}
\centering
\begin{tabular}{llcccc}
\hline
\textbf{Group} & \textbf{Task} & \makecell{\textbf{Code Review} \\(Within-sub.)}  & \makecell{\textbf{Debugging}\\(Within-sub.)} & \makecell{\textbf{Refactoring}\\(Within-sub.)}   & \makecell{\textbf{Implement new features}\\(Within-sub.)} \\ 
\hline

Control & T0 vs. T1 &  \makecell{p = 0.537}  & \makecell{p = 0.475} & \makecell{p = 0.519} & \makecell{p = 0.753}\\
\hline

Control & T0 vs. T2 &  \makecell{p = 0.887} & \makecell{p = 0.283} & \makecell{p = 0.548}  & \makecell{p = 0.287}\\
\hline
 
Control & T1 vs. T2 &  \makecell{p = 0.786}  & \makecell{p = 0.816} & \makecell{p = 0.859}  & \makecell{p = 0.094}\\
\hline

Experiment & T0 vs. T1 &  \makecell{p = 0.531}  & \makecell{p = 0.219} & \makecell{p = 0.085}  & \makecell{p = 0.842}\\
\hline

Experiment & T0 vs. T2 &  \makecell{p = 0.536}  & \makecell{p = 0.710}  & \makecell{p = 0.977}  & \makecell{p = 0.470}\\
\hline
 
Experiment & T1 vs. T2 &  \makecell{p = 0.127}  & \makecell{p = 0.013} & \makecell{p = 0.009} & \makecell{p = 0.157}\\
\hline

\end{tabular}
% \vspace{0.2em}
\end{table}

We apply statistical tests to the paired tasks for each group and on each SE activity for within-subjects analyses. Looking at Table~\ref{tab:learning-confidence-with-subject}, we can see there are seemingly significant differences in the experiment group between Task 1 and Task 2 for debugging and refactoring. This echoes the confidence declines that we observed in Figure~\ref{fig:group-confidence}, suggesting that the subsequent removal of the tool GazePrinter caused these consequences in the following task. However, since we are comparing the same group data in different aspects/activities, a multiple comparisons problem~\cite{wiki2026multicomparisonsprob} needs to be treated. For this purpose, we apply the Bonferroni correction~\cite{wiki2026bonferronicorrection}, which leads to a lower p-value threshold (decreasing the threshold from 0.05 to 0.002), to double-check the statistical significance. After this correction, no statistical significance is observed.  

In summary, we can speculate that the experiment group's confidence increased in Task 1 with the presence of our tool, GazePrinter. This is evidenced by the confidence changes between Task 0 (pre-experiment) and Task 1, particularly in debugging and refactoring. However, this increase did not sustain in Task 2, wherein the tool support was removed. Conversely, the confidence of the experiment group even dropped slightly below their initial confidence at the beginning of the study. This perhaps indicates a double-edged sword effect, partly echoed by Participant P0307S2EC's quote from our interview results. 

\textbf{Perceived Understanding.}~We examine the perceived understanding of the codebases by participants. In connection with that, we also look into the readability metric reported by them to make an inference on whether it could have affected their understanding, especially if there is a significant difference between the two groups. 

\begin{table}[h]
\caption{Summary of perceived readability and understanding (max=10) (* p < 0.01 with Bonferroni correction, 0.05/4).}
\label{tab:learning_understanding_stats}
\centering
\begin{tabular}{llcccc}
\hline
\textbf{Task} & \textbf{Group} & \textbf{Readability} & \textbf{Between-subjects} & \textbf{Understanding of the Codebase} & \textbf{Between-subjects}\\
\hline
Task1 & Control & 6.16 (\textpm 2.12) & \multirow{2}{*}{p-value = 0.690} & 3.89 (\textpm 1.94) & \multirow{2}{*}{p-value = 0.063}\\
 & Experiment & 6.37 (\textpm 1.77) & & 5.05 (\textpm 1.84) &\\
\hline
Task2 & Control & 5.42 (\textpm 2.17) & \multirow{2}{*}{p-value = 0.813} & 3.89 (\textpm 2.31) & \multirow{2}{*}{p-value = 0.280}\\
 & Experiment & 5.21 (\textpm 1.90) & & 4.63 (\textpm 2.17) &\\
 \hline
\end{tabular}
\end{table}

As shown in Table~\ref{tab:learning_understanding_stats}, the readability of the codebase is comparable between the two groups and shows no statistical difference. This is the case for both Task 1 and Task 2. However, the difference in perceived understanding between the two groups is noteworthy; there is a discrepancy of 1.16 in Task 1 and 0.74 in Task 2. Though both differences are not statistically significant, the p-value of Task 1 is close to 0.05, implying a much larger difference between the two groups. Comparing each group's perceived understanding between the two tasks, the control group has no change, and the experiment group experiences a slight decrease of 8\%. To sum up, our tool GazePrinter perhaps enhanced participants' perceived understanding of the codebase. The specific degrees of understanding both groups perceived are not on the high end (e.g., larger than 6) and perhaps were affected by the code readability per se, but we can at least rule out that the difference they demonstrated in perceived understanding is not because the readability of codebases was significantly different to them. 

\textbf{Perceived Helpfulness.}~To gauge learning experience, we also ask participants to indicate to what extent they believe Task 1 helped them in Task 2. This metric discloses participants' perception of the transferable knowledge they gained, as well as surfaces a subjective factor that potentially compounds the learning. 

\begin{table}[h]
\caption{Summary of perceived helpfulness (max=10) per group.}
\label{tab:learning_helpfulness_stats}
\centering
\begin{tabular}{lcc}
\hline
\textbf{Group} & \textbf{Helpfulness} (Task 1 helped Task 2) & \textbf{Between-subjects}\\
\hline
Control & 5.74 (\textpm{2.49}) & \multirow{2}{*}{p-value = 0.498}\\
Experiment & 6.21 (\textpm{2.46}) & \\
\hline
\end{tabular}
\end{table}

We can see in Table~\ref{tab:learning_helpfulness_stats} that the experiment group perceived an 8\% higher degree of helpfulness from Task 1, with a difference of 0.47. However, this difference is not statistically significant, as indicated by the p-value of 0.498. On average, both groups reported that an intermediate level of learning occurred between the tasks. 

\vspace{2mm}
\setlength{\fboxsep}{7pt}%box to content distance
\setlength{\fboxrule}{2pt}%thickness of box
\fcolorbox{gray!60}{gray!20}{%
    \parbox{0.88\columnwidth}{%
        \textbf{Finding 6}: The experiment group reported higher confidence and perceived understanding when the tool GazePrinter was present. Both groups acknowledged an intermediate degree of learning between tasks. 
        }
}
\vspace{2mm}

\subsubsection{Interview Results: Learning}
%
% Ebbinghaus’ Forgetting Curve
% TODO: don't forget about this comment
%
%\todo[author=Peng]{perhaps add quotes from participants}

We mainly report the immediate learning, but also mention potential long-lasting learning outcomes. According to Table \ref{table:interview-gp}, 82.5\% of the participants recognized that reading and understanding the first codebase helped them in reading and understanding the next one. This implies there was an obvious, immediate learning effect between the two tasks. Most participants reported that working with the first codebase either helped refresh their memories of the Java programming language syntax or simply helped warm up their brains into a state suitable for programming, or both (see quote from Participant P0317S1ED). Some attributed the reason to the similar architectures (both are web applications with an MVC structure) and the technical stacks (both use the Java Spring Boot framework). A few mentioned that it was because they had already become aware of what was expected for the hands-on part of the task and knew roughly where to look at or make changes in the second codebase. 

% \vspace{2mm}
% \vspace{2mm}
\begin{figure}[H]
\centering
\setlength{\fboxsep}{7pt}%box to content distance
\setlength{\fboxrule}{2pt}%thickness of box
\fcolorbox{gray!30}{gray!10}{%
    \parbox{0.9\columnwidth}{%
        P0317S1ED. \textit{``...Yeah, it definitely helped me get back into the web development thinking, because I haven't done it in so long and [...] it kind of warmed up my brain for the second codebase.''}
        }
}
\label{box:quote-learning}
\end{figure}
% \vspace{2mm}

This corroborates that programmers welcome and are good at capturing recurring patterns~\cite{jbara2017experiment, maalej2014developercomprehension}. Some participants even commented that participating in the experiment helped them learn more about programming in general, since they normally work with another language, such as Python, and code for other purposes, e.g., data analysis. It is unclear how long such unplanned learning can be sustained in participants' memories and in what concrete way it may help them with future programming activities. But this perhaps implies that even for specialized programming education or orientation targeting newcomers, it is still necessary to account for a reasonable degree of variety in introducing technical tasks and stacks~\cite{happe2021multidisciplinarity}. 
% breadth and T-shape depth
% deep specialists and expert generalists (reference)

\begin{table}[h]
\caption{Summary of interview results regarding learning (Task 1 helped Task 2) per group.}
\label{tab:learning_interview_stats}
\centering
\begin{tabular}{lcccccc}
\hline
\textbf{Group} & \textbf{Positive} & \textbf{Negative} & \textbf{Odds} (positive) & \makecell{\textbf{Odds Ratio} \\(between-subjects)} & \textbf{p-value} & \textbf{Confidence Interval}\\
\hline
Control & 18 & 1 & 18.00 & \multirow{2}{*}{4.8} & \multirow{2}{*}{p = 0.340} & \multirow{2}{*}{[0.483, 47.684]}\\
Experiment & 15 & 4 & 3.75 & & &\\
\hline
\end{tabular}
\end{table}

We further examine the distribution of positivity among the two groups and whether a statistically significant difference exists between the two groups. As shown in Table~\ref{tab:learning_interview_stats}, the control group is 4.8 times more likely to report a positive learning effect from Task 1 than the experiment group. However, their difference in this respect is not significant as the p-value suggests. In other words, this difference can be attributed to randomness in the two groups. Since the confidence interval is wide and contains the value of 1, we are unsure about the difference, and there is a possibility that the two groups bear no difference. 

\vspace{2mm}
\setlength{\fboxsep}{7pt}%box to content distance
\setlength{\fboxrule}{2pt}%thickness of box
\fcolorbox{gray!60}{gray!20}{%
    \parbox{0.88\columnwidth}{%
        \textbf{Finding 7}: Both groups reported an immediate learning effect from Task 1. Task 1 helped participants warm up their brains and recall the syntax of the programming languages and technical stack. 
        }
}
\vspace{2mm}

\subsection{Experience of Visualization of Expert Gaze (\RQ{4})}
%
% \todo{Peng: statistical significance between control and experiment for learning from the interview - done by Peng}

Here, we share participants' user experience with regard to the usefulness and usability of the visualization of expert gaze via GazePrinter. The first part of the results is based on the questionnaires with participants from the experiment group (N=19) and a pilot study (N=1). The second part of the results is based on the interviews with all the participants. The metrics derived from the questionnaires are presented in Table~\ref{tab:gp_ux_stats}, and the results gathered from the interviews are summarized in Table~\ref{table:interview-gp}. 

{
\setlength{\tabcolsep}{2pt} % default is 6pt
\begin{table}[htbp]
\caption{Summary of interview results regarding experience of GazePrinter, participant background, experience of of the experiment, and learning between tasks.}
\label{table:interview-gp}
\centering
% \begin{tabular}{|l|c|c|c|l|l|}
\resizebox{\textwidth}{!}{%
\begin{tabular}{l c c c l l}
\hline
\textbf{Theme} & \textbf{Topic} & \textbf{Category} & \textbf{Code} & \textbf{Count} & \textbf{Percentage}\\
\hline

\multirow{10}{*}{\makecell[l]{GazePrinter\\ \footnotesize(Experiment group, N=19; \\\footnotesize Pilot study, N=1; total N=20)}} & \multirow{3}{*}{Usefulness} & \multirow{3}{*}{Initial usage} & Positive & 16 & \textcolor{green!70!black}{80\%}\\
\cline{4-6}
&  & & Mixed & 3 & 15\%\\
\cline{4-6}
&  & & Negative & 1 & 5\%\\
% \cline{4-6}
% &  & & Not relevant & 20 & \\
\cline{2-6}
& \multirow{3}{*}{Usability} & \multirow{3}{*}{Initial usage} & Expressively positive & 4 & 20\%\\
\cline{4-6}
&  & & Non-negative & 14 & \textcolor{green!70!black}{70\%}\\
\cline{4-6}
&  &  & Negative & 2 & 10\%\\
\cline{2-6}
& \multirow{4}{*}{Suggestions} & \multirow{8}{*}{Application Scenario} & general collaboration & 2 & 10\% \\
\cline{4-6}
&  &  & code review & 2 & 10\% \\
\cline{4-6}
&  &  & debugging & 1 & 5\% \\
\cline{4-6}
&  &  & filtering & 1 & 5\%\\
\cline{4-6}
&  &  & open-source project & 3 & 15\% \\
\cline{4-6}
&  &  & large files & 2 & 10\% \\
\cline{4-6}
&  &  & legacy code & 1 & 5\%\\
\cline{4-6}
&  &  & documentation & 1 & 5\%\\
\cline{3-6}
&  & \multirow{3}{*}{Highlighting Mechanism} & prefer current design & 4 & 20\%\\
\cline{4-6}
&  & & prefer highlighting code  & 7 & 35\%\\
\cline{4-6}
&  & & prefer having both  & 1 & 5\%\\
\cline{3-6}
&  & Color \& Layout & custom colors, intensity, etc. & 3 & 15\%\\
\cline{3-6}
&  & New features & textual tags, etc. & 5 & 25\%\\
\hline

\multirow{6}{*}{Background} & \multirow{3}{*}{IDE} & \multirow{3}{*}{Impact} & Positive & 6 & 15\%\\
\cline{4-6}
 &  &  & Neutral & 26 & \textbf{65\%} \\
\cline{4-6}
 &  &  & Negative & 8 & 20\% \\
\cline{2-6}
 & \multirow{3}{*}{Framework} & \multirow{3}{*}{Bootstrap} & No experience & 28 & \textbf{70\%} \\
\cline{4-6}
 &  &  & With Experience & 8  & 20\%\\
 \cline{4-6}
 &  &  & Not asked & 4  & 10\%\\
\hline

\multirow{9}{*}{Experiment} & \multirow{11}{*}{Perceived Complexity} & \multirow{4}{*}{Task} & Task 1 > Task 2 & 12 & 30\%\\
\cline{4-6}
& & & Task 1 < Task 2 & 18 & 45\%\\
\cline{4-6}
& & & Task 1 == Task 2 & 3 & 7.5\%\\
\cline{4-6}
& & & Not asked & 7 & 17.5\%\\
\cline{3-6}
& & \multirow{6}{*}{Reason} & \makecell{Availability of online tools \\\footnotesize(e.g., ChatGPT, Google)} & 5 & 12.5\%\\
\cline{4-6}
& &  & Locating code & 15 & 37.5\%\\
\cline{4-6}
& & & Unfamiliarity & 25 & \textcolor{red!80!black}{62.5\%}\\
\cline{4-6}
& & & Structure & 11 & 27.5\%\\
\cline{4-6}
& &  & Volume & 7 & 17.5\%\\
\cline{4-6}
& &  & Absence of GP & 4 & 10\%\\
\cline{4-6}
& &  & Other & 15 & 37.5\%\\
\hline
\multirow{3}{*}{Learning} & \multirow{3}{*}{Pattern Re-application} & \multirow{3}{*}{\makecell{Within Session \\ \footnotesize(Task 1 helped Task 2)}} & Positive & 33 & \textcolor{green!70!black}{82.5\%} \\
\cline{4-6}
& & & Negative & 5 & 12.5\% \\
\cline{4-6}
& & & Not asked & 2 & 5\% \\
\hline
\end{tabular}
}

\end{table}
}

\textbf{Questionnaire Results.}~For usefulness, the experiment group reported a score of 5.45 on average. Under this category, they reported a mean score of 6.15 in response to whether the tool GazePrinter helped them understand the codebase more quickly, and 4.40 in response to whether it helped them understand it better. For usability, they reported a mean score of 5.95. All the scores are on a scale of 10, with 10 indicating the maximum positivity. To sum up, participants who were exposed to GazePrinter recognize both its usefulness and usability on the intermediate level. Participants expressed a more positive view of the tool's usability than its usefulness. In terms of usefulness, participants perceived that it was more useful for understanding the codebase more quickly than for understanding the codebase better. 

% \todo{Peng: move table 18 to section 5.5 about GazePrinter experience}

\begin{table}[H]
\caption{Summary of user experience with GazePrinter (max=10) reported by the experiment group.}
\label{tab:gp_ux_stats}
\centering
\begin{tabular}{lcccc}
\hline
\textbf{Group} & \textbf{Usefulness} & \textbf{a) Understand quicker} & \textbf{b) Understand better} & \textbf{Usability} \\
\hline
Experiment & 
5.45 (\textpm{2.21}) &
6.15 (\textpm{2.62})&
4.40 (\textpm{2.30})& 
5.95 (\textpm{2.28})\\
\hline
\end{tabular}
\end{table}

% +pilot
% \begin{table}[h]
% \caption{GazePrinter User Experience (max=10).}
% \label{tab:gp_ux_stats}
% \centering
% \begin{tabular}{lcccc}
% \hline
% Group & Usefulness & a) Understand quicker & b) Understand better & Usability \\
% \hline
% Experiment & 
% 5.45 (\textpm{2.21}) &
% 6.15 (\textpm{2.62})&
% 4.40 (\textpm{2.30})& 
% 5.95 (\textpm{2.28})\\
% \hline
% \end{tabular}
% \end{table}

% \todo{Peng: review integration of Table 22 and formulation of Finding 8 - done by Peng}

\textbf{Interview Results.}~For usefulness, 80\% of the participants gave positive feedback, 15\% had mixed perceptions about it, and 5\% did not find it so. Participants had mixed feelings because they believed the tool would be useful for other programmers, although not so useful for them during the experiment. Many of them also deemed that having such a type and layer of support integrated into the code would help them feel more confident and comfortable working with an unseen codebase. One participant who indicated not useful was bothered as the tool gave no explanation why certain parts of the code were more important, and some code blocks were equally important in her view. Nonetheless, the participant still believed the tool would facilitate locating critical files and code more quickly, and deciding where to focus, and increase one's confidence and comfort working with a new codebase. 
%Noticiably, the particpant who rated usefulness negatively actually admit she started to appreciate the tool when she had to work with the second codebase without any visual assistance like what the tool provided in the first codebase. 
Another participant actually began to appreciate the tool when it was absent in the second codebase, though the initial perception of the tool was of low usefulness (see quote from Participant P0307S2EC).  

% \vspace{2mm}
\begin{figure}[H]
\centering
\setlength{\fboxsep}{7pt}%box to content distance
\setlength{\fboxrule}{2pt}%thickness of box
\fcolorbox{gray!30}{gray!10}{%
    \parbox{0.9\columnwidth}{%
        P0307S2EC. \textit{``...But after like the second time, I didn't have it. OK, it looks so weird now. Because [...] in the past, it's like several blocks, so I can ... just like differentiate between each block[s]. But now, even though I have the blanks, it's just like everything is together, and it's hard to find. So after that, I find, OK, it's actually useful. ''}
        }
}
\label{box:quote-useful}
\end{figure}
% \vspace{2mm}

For usability, 90\% of the participants stated that the design, including the choice of layout, shape, and colors, was acceptable and non-disturbing. Among these participants, four participants, who constitute 20\% of the total experiment group, expressively complimented the tool design. 10\% of the participants complained that there was room for improvement. The complaints included that it required effort to look back and forth between the heat bars and the code blocks, and it might compete with pre-existing mechanisms for the communication channel, which is the gutter next to the code in the editor.

We asked participants to consider other potential user scenarios and give preference between the current highlighting mechanism (heat bar in the gutter) and a conventional alternative (directly highlighting the code/text). In addition to suggestions, some participants also voluntarily shared inputs for other aspects of the tool, such as the color, layout, and possible new features. In terms of potential user scenarios, most participants mentioned that the tool or this kind of gaze-based support could be useful in collaborative settings among programmers within an organization. However, the intents or concrete ways of making use of such visual aids varied or switched among these participants. Some wanted to pay more attention to where their collaborators looked, e.g., for locating potential causes during debugging, whereas some wanted to look at areas their collaborators ignored or neglected, e.g., for code review, so as to reach better code coverage. This is in line with what we have found in our previous design study~\cite{kuang2024designing}. Some participants suggested that visualization of expert gaze would be useful for navigating and understanding open-source codebases for the first time, or for those programmers who were less experienced, helping them skip less relevant files and thus reducing waste of time and confusion. This echoes our motivation for conducting this study. A participant even envisioned it could be applied to documentation, for instance, for a software library. Other suggestions included calibrating the accuracy of the mapping between the colors and criticalness of the code fragments, making the current color tone more striking, providing more categories of colors as options and allowing for adding tags associated with them, displaying related diagrams when hovering on the highlighted code fragments, and explicitly communicating the reason why the highlighted code fragments were important. 

\vspace{2mm}
\setlength{\fboxsep}{7pt}%box to content distance
\setlength{\fboxrule}{2pt}%thickness of box
\fcolorbox{gray!60}{gray!20}{%
    \parbox{0.88\columnwidth}{%
        \textbf{Finding 8}: The majority of the participants (above 80\%) in the experiment group gave positive feedback on the usefulness and usability of our designed tool assistance. 
        }
}
\vspace{2mm}

%\vspace{2mm}
%\setlength{\fboxsep}{7pt}%box to content distance
%\setlength{\fboxrule}{2pt}%thickness of box
%\fcolorbox{gray!60}{gray!20}{%
%    \parbox{0.88\columnwidth}{%
%        \textbf{Result summary 5}: The Majority of the participants in the experiment %group gave positive feedback on the usefulness and usability of our designed tool assistance. 
%        }
%}
%\vspace{2mm}

%Additionally, for certain yes-or-no-like questions, we also collected data about them in either the pre-experiment or the post-experiment questionnaire, so they were given lower priority when we had to select from competing questions. 

%\vspace{2mm}
%\setlength{\fboxsep}{7pt}%box to content distance
%\setlength{\fboxrule}{2pt}%thickness of box
%\fcolorbox{gray!60}{gray!20}{%
%    \parbox{0.88\columnwidth}{%
%        \textbf{Result summary 4}: The Majority of the participants reported a top-down or integrated strategy for navigating a codebase for the first time. AI has become the primary tool to which they sought help for understanding unfamiliar code. The time required for participants to understand unseen code varies to a large extent, ranging from a year to minutes, depending on the scale of the code involved. However, most of them indicated a time of hours. In terms of experience, the majority reported neutral or mixed sentiment with codebase comprehension tasks. 
%        }
%}
%\vspace{2mm}

\subsection{Participants' Perceived Complexity of the Experiment and its Causes}

We report participants' perception of the relative complexity of the two codebases and the reasons causing it, listed in Table~\ref{table:interview-gp}. Among the participants, 45\% stated that the second codebase was more difficult to understand than the first codebase, while 30\% indicated the opposite. Three participants (7.5\%) reported that the two codebases were equally difficult for them. 

In terms of reasons leading to that perception, the most outstanding one was related to unfamiliarity (62.5\%), for instance, not proficient enough in the programming language Java, little or no exposure to the frontend software technical stack, such as HTML, CSS, JavaScript, or having limited experience with programming in general. The second most common reason was locating relevant code, which might be a manifestation of the previously mentioned unfamiliarity. Because participants did not understand the code, they were unable to identify which specific part of the code was relevant to the task at hand. Other leading causes mentioned included the structure, volume, and other characteristics, such as the naming convention, of the designated codebase. In addition, some reported that the availability of access to ChatGPT and Google caused them to perceive the second codebase as more difficult. Similarly, several participants from the experiment group who had been introduced to the tool GazePrinter mentioned that it was in part because of its absence in the second codebase.

%\vspace{2mm}
%\setlength{\fboxsep}{7pt}%box to content distance
%\setlength{\fboxrule}{2pt}%thickness of box
%\fcolorbox{gray!60}{gray!20}{%
%    \parbox{0.88\columnwidth}{%
%        \textbf{Result summary 7}: Both groups reported that there was an immediate learning effect from Task 1. Task 1 helped participants warm up their brains as well as recall the syntax of the programming languages and the technical stack. 
%        }
%}
%\vspace{2mm}

\section{Discussion}
\label{sec:discussion}

% \todo{incorporate links to findings to support the answers to the RQs, for instance like " ... (Finding 1) ..." - done by Peng}
We consolidate the results to answer our RQs and discuss their implications. 
\begin{itemize}
    % \todo{Peng: reduce text and update references to findings - done by Peng}
    
    \item \RQ{1}: The intervention of our tool did not render a statistically significant difference between the two groups in efficiency, effectiveness, and cognitive effort, which are measured through response time, task performance, and cognitive load, respectively. That said, we observe weak evidence indicating a reduction in response time and cognitive load in the subsequent task for the experimental group (Finding 1). 

    \item \RQ{2}: In terms of reading strategies, the two groups exhibited differences in the file and module reading orders (Finding 2 \& 3), as well as in the file- and line-level attention distributions (Finding 4 \& 5). On the file level, the two groups demonstrate statistically significantly different reading orders (Finding 2). On the module level, the experiment group reads more consistently across tasks (Finding 3). In terms of attention distribution, the experiment group aligns more closely with the experts (Finding 4). On the line level, the experiment group also looks more similar to the experts (Finding 5).  
    
    % we found that our tool did not significantly reduce the time the experiment group spent reading the codebases (Finding 1). Although the experiment group spent less time reading the second codebase, they spent slightly more time on the first one.  The experiment group did not outperform their counterpart in completing the tasks. They received slightly fewer points for both tasks; however, the differences between the two groups are not significant. The experiment group reported a lower cognitive load for both tasks (Finding 3). But the cognitive load differences again are not significant between the two groups. We found that the experiment group read fewer files (Finding 5), they exhibited a significantly different file reading order from the control group (Finding 2), and their gaze patterns aligned more closely with the experts (Finding 3 \& 4). 
    
    \item \RQ{3}: Both groups reported an intermediate level of immediate learning from Task 1 to Task 2 (Finding 6 \& 7). Perceptually, the experiment group experienced a higher level of confidence, perceived understanding, and perceived helpfulness compared to the control group (Finding 6), though none of the differences is statistically significant. Physiologically, this learning effect manifests in helping them activate their brain function and retrieve prior knowledge from memory (Finding 7).
    % Additionally, the experiment group reported more positively in perceived understanding of the codebases and perceived helpfulness of Task 1. They also disclose a larger increase in confidence when our tool is present.

    % \todo{Peng: add the answer to RQ3- done by Peng}
    \item \RQ{4}: The majority of participants who used GazePrinter in the experiment reported positively about its usefulness and usability~(Finding 8). In particular, they perceive the tool to be more useful in helping them understand the codebase faster than in helping them understand it better. 
    
\end{itemize}
 %This is in part attributed to the absence of our tool in Task 2, which creates a sense of a lack of support, according to what was reported from the experiment group. 
Overall, our findings suggested that the integration of gaze-based assistance into the codebase had a stronger impact on programmers' self-assessed mental load, perceptions, and reading patterns in terms of reading order and attention distribution. We assume that these variables play an important role in characterizing programmers' program comprehension strategy. 

\subsection{Gaze-assisted Communication Between Programmers}
Gaze plays an important role in human communication~\cite{bailly2010gaze, degutyte2021role}. While its link to programmers' attention has been well studied in the software engineering context, its potential to aid communication between programmers is less explored. Our tool visualizes the shared gazes of a codebase in a programming environment to assist programmer communication, integrating the design into three channels/mechanisms that the environment has already been utilizing to communicate with programmers. Previous research reports that development teams often experience communication frictions~\cite{defranco2017communication}, and there is also a gap between software engineering practice and its education~\cite{oguz2019edugap}. The former is particularly prominent among distributed development teams and for asynchronous \gls{SE} activities such as code review~\cite{sadowski2018codereview}. The latter stressed the importance of educating students about software maintenance, which demands proper training on program comprehension and communication between programmers about the shared codebase. Contextualized in a distributed setting that underscores the necessity of asynchronous communication, our study shows that it is viable to nudge the code-reading behaviour of novices to better align with experts within a pre-existing codebase by strategically placing gaze-derived visual cues in the programming environment. It shows that designing gaze-based assistance into the code space is of great potential in supporting the onboarding of newcomers. Such an exploration further paves the way for developing gaze-aware programming environments or AI agents to collaborate with programmers in the era of software 3.0~\cite{li2025softengv3}. 

\subsection{Revisiting Code Complexity in the Context of Program Comprehension}
% relate to the literature in program comprehension

% \todo{Peng: connect the presentation in Fig 10 to the results from the study and observations made in relation to the program comprehension - done by Peng}

Code complexity is a variable that often comes up when researchers and programmers discuss program comprehension. Here, we reflect on the concept and approximation of code complexity in relation to program comprehension by integrating insights gained from the results of this study, both the quantitative and qualitative ones. Based on the perceptual measures (e.g., perceived readability, understanding, and helpfulness) we gathered and the perception-oriented probing questions (e.g., perceived causes of complexity) during the interviews, we argue for the consideration of programmers' perception as a contributor to code complexity and attempt an articulation of it. To illustrate our reasoning, we draw its relationships and interactions with other key variables that we examined in this study, focusing on the code-reading slice of the program comprehension process (as shown in Figure~\ref{fig:code-reading-modeling}). 

Code complexity underlies the cognitive effort involved during program comprehension, making it worthy of a second look with an additional physiological lens. Among many approaches to capture it, cyclomatic complexity~\cite{mccabe1976complexity} (a.k.a. McCabe complexity) is the one that has been adopted most widely. In addition to this, another approximation that came into our attention is the so-called cognitive complexity introduced by SonarSource~\cite{campbell2018cogcomplexity}\footnote{We acknowledge that cognitive complexity may be a general term in other disciplines such as cognitive neuroscience. Here, we use it as a specific reference to the work by Campbell et al.~\cite{campbell2018cogcomplexity}}, which originally aimed to aid code understandability and ultimately maintainability. It was further empirically inspected by Munoz et al.~\cite{munoz2020cogcomplexity} on its correlation to a range of variables, including comprehension time and correctness, perceived difficulty, physiological metrics, and code understandability. Cognitive complexity is plausible in reminding us of the shortcomings of cyclomatic complexity. It pinpoints the relative nature of the same code to different programmers and shows a viable way to factor in this variability in perception to produce a more reliable approximation for code complexity. 

\subsubsection{Cognitive Complexity and Cyclomatic Complexity}

More concretely, cognitive complexity attempts to accommodate code's relative understandability to approximate code complexity more accurately than cyclomatic complexity by including modern language constructs and applying aggregation instead of naive increment~\cite{campbell2018cogcomplexity}. For example, for a Java switch statement comprising five cases, cyclomatic complexity increases by 1 for each case, ending up with a total of 5; cognitive complexity counts all the cases together as 1, ending up with a total of 1. However, methodologically, cognitive complexity still looks at code only and calculates its understandability mathematically. Hence, it carries the same drawback as cyclomatic complexity (though arguably a better approximation). That is, it did not paint the full picture of code complexity as it does not recognize or articulate the perception of complexity, and how it can be influenced by the programmer's fluid proficiency~\cite{kuang23gazedevtool}. For instance, even for the same programmer, the understandability of the same nested method will appear different the second time from the very first time (another effect acknowledged by our participants as shown in Table~\ref{table:interview-pc}, under the 'Source of Difficulty' category). However, using the algorithm of either cognitive complexity or cyclomatic complexity, its understandability remains constant for the programmer, even at different time points. Another simple example that can challenge both measures is that none of them can capture the complexity or understandability caused by inconsistent code style and over-abbreviated variable names~\cite{maalej2014developercomprehension}.

% perceptual fluency, visual expertise, cognitive walkthrough

\subsubsection{Perceived Complexity}

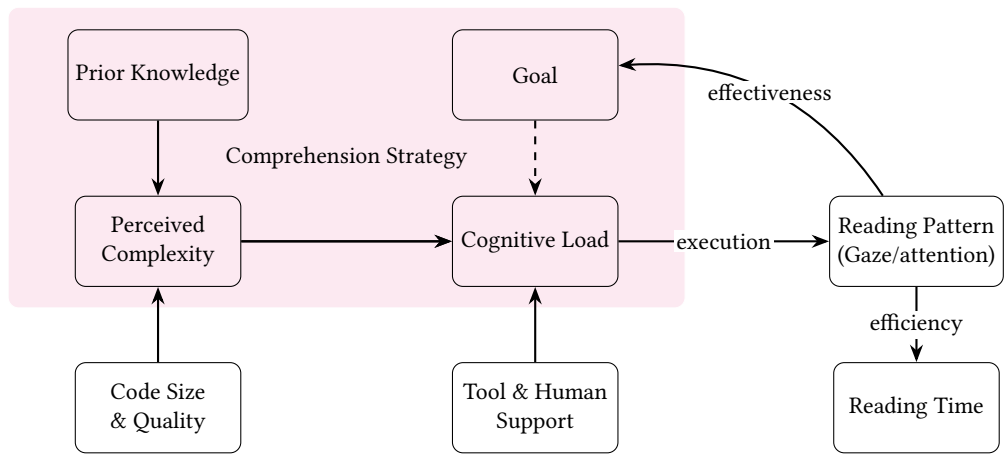
\begin{figure}[htbp]
\centering
\begin{tikzpicture}[
  node distance=1cm and 2.8cm,
  box/.style={draw, rounded corners, minimum width=2.2cm, minimum height=1.2cm, align=center},
  arrow/.style={-{Stealth}, thick},
  dashedarrow/.style={-{Stealth}, thick, dashed}
]

% Nodes
\node[box] (cl) {Cognitive Load};
\node[box, right=of cl] (rp) {Reading Pattern \\(Gaze/attention)};
\node[box, below=of rp] (rt) {Reading Time};
\node[box, left=of cl] (cm) {Perceived \\Complexity};
\node[box, above=of cm] (pk) {Prior Knowledge};
\node[box, below=of cm] (csq) {Code Size \\\& Quality};
\node[box, above=of cl] (gl) {Goal};
\node[box, below=of cl] (ths) {Tool \& Human \\Support};

% Layer setup
\pgfdeclarelayer{background}
\pgfsetlayers{background,main}

% Inside tikzpicture
\begin{pgfonlayer}{background}
  \node[draw=white, fill=magenta!10, rounded corners, 
        minimum width=9cm, minimum height=4cm] 
        at ($(pk)!0.5!(gl)+(0,-1.1)$) {Comprehension Strategy};
\end{pgfonlayer}
%formulation and update

% Arrows
\draw[arrow] (pk) -- (cm);
\draw[arrow] (cm) -- (cl);
\draw[arrow] (csq) -- (cm);
\draw[arrow] (cm) -- (cl);
% \draw[arrow] (cl) -- (rp);
% \draw[arrow] (rp) -- (rt);
\draw[dashedarrow] (gl) -- (cl);
\draw[arrow] (ths) -- (cl);

\draw[arrow]
  (cl) -- 
  node[midway, fill=white, inner sep=1pt] {execution}
  (rp);
  
\draw[arrow]
  (rp) to[bend right=30]
  node[midway, fill=white, inner sep=1pt] {effectiveness}
  (gl);

\draw[arrow]
  (rp) -- 
  node[midway, fill=white, inner sep=1pt] {efficiency}
  (rt);
  
\end{tikzpicture}
\caption{Comprehension strategy formulation and execution (during code reading).}
\label{fig:code-reading-modeling}
\end{figure}

In Figure~\ref{fig:code-reading-modeling}, we map out our understanding of what constitutes and modulates code complexity by slicing program comprehension at code reading for a close-up inspection. We posit that complexity has two dimensions. While there is a static aspect embedded in the code (e.g., attributes such as size and quality), it is more a dynamic perception that resides in a programmer's mind, influenced by their constantly updating prior knowledge. This is in part supported by our interview results, where the participants report unfamiliarity (the lack of prior knowledge) as the primary cause for complexity (as shown in Table~\ref{table:interview-gp}, under the category 'Perceived Complexity - Reason'). We term this notion as perceived complexity. During code reading, this perceived complexity will predominantly determine the degree of cognitive load incurred. However, external support, such as from another programmer or a tool, will help modulate the mental load experienced by the programmer. Internally, the goal the programmer embodies at the time of reading also affects the cognitive load. It motivates the programmer to curate different degrees and ranges of concentration, which then manipulates the cognitive load. The fluctuation of perceived complexity and cognitive load will drive the programmer to formulate and execute their comprehension strategy, exhibited as patterns in their reading actions. Eventually, these actions lead to different time costs for the programmer. 

\section{Contribution}
% relate to the literature in SE eye tracking study or contribution to the knowledge in the studied domain
In summary, our work contributes to the knowledge of source code or program comprehension in three aspects. First, our study is the first to use eye-tracking to examine program comprehension in the context of onboarding new codebases at the scale of involving hundreds of files and thousands of lines of code. This piece of work bridges previous studies of single-file code snippets with common industrial practice on codebases with tens of thousands of lines of code and above. Second, we adopted a mixed-methods approach to gather both physiological and self-reported data to triangulate findings with respect to how gaze-based assistance can affect programmers in their acting and perception of codebase comprehension. Our findings both corroborate and extend the research community's knowledge about programmers' formulation and execution of program comprehension strategies, as well as capture some modern and emerging strategies%
% and nuanced variability among novices
%, such as action-assisted and AI-tool-assisted comprehension
. Third, our modeling of the cognitive process of program comprehension, as observed in a slice of code reading, provides an alternative explanation for the interactions between the key variables involved in this procedure. This may help to create a shared understanding of program comprehension for novice programmers among researchers who use eye tracking to conduct empirical software engineering studies with this population. 

% novice programmers' cognitive model of code comprehension
% proficiency of programming languages is relative and fluid
% so what can influence proficiency/complexity
% motivation of compensation with vs. without
% goal - scope, volume of code needed to read
% task-driven reading
% walkthrough/tour/orientation, alike reading

% \input{figs/two-perspectives}
%limitation to program with a GUI
% implication for programming education:
%- breadth (exposure) vs depth (T-shape)
% new cognitive model for PC
%shift right, application of knowledge gained from PC research

% future work
\subsection{Future Work}
We claim that program comprehension becomes even more important in the era of AI. This is because both the amount of code and the novice programmer population are increasing due to the ease of producing code using AI. However, AI can also introduce vulnerabilities and other types of defects into the code without notifying programmers. Studies show that developers still hold their trust in AI-generated code to some extent~\cite{cheng2024trust}. This can be even more challenging for novices with less programming experience. Hence, there remains a solid need for programmers to invest time and mental effort to audit perhaps an even larger amount of code than before. In contrast, programmers' attention and cognitive capacity are limited; at least, in the near term, they cannot increase in accordance with the speed of AI. To this end, understanding how programmers perform source code comprehension at scale becomes even more relevant. Future work can explore how programmers' visual attention can be sensed and interpreted by intelligent interactive programming systems. Researchers can also test with professional developers or a different cluster of novice programmers, such as designers and educators, to verify if the support provided by gaze-based visual assistance generalizes to other groups of programmers and/or other software engineering tasks.

\section{Threats to Validity}
\label{sec:threats}

% \todo{Peng: Add something about Bonferroni here? - Peng: my understanding is that once we have corrected it and there is no statistical significance reported, then no risk was introduced in this regard? Should we talk about the monetary incentives, though? We also seemed to forget to mention this in the method}

% \todo{Peng: discuss open and closed exams to task performance here and in results}

We evaluate the threats to the construct, internal, and external validity of our study.

\textbf{Construct Validity.}~The time prescribed in our experimental design for code reading is relatively short for working on a codebase with thousands of lines of code. In real-world practices, programmers may take a much longer time. However, there is also a possibility that programmers may reach fatigue if we completely imitate the real-world reading. Furthermore, some programmers' reading is likely to be more sporadically distributed among multiple turns within a period of time. In that case, we would not be able to control the variables of the experiment enough to perform meaningful comparisons and draw conclusions. The way we divide code reading and making code changes into two separate sessions helps avoid an overly complicated experimental design and minimizes possible fatigue while keeping it representative as well as comparable with common practices in the literature. The time intervals we designed to test participants' understanding also fit into the timeframe in which learners retain newly acquired knowledge most effectively~\cite{murre2015forgettingcurve}. 

Another threat is that participants performed code reading only with a high-level goal (try to understand the codebase as much as possible to establish an overview), but without knowing the specific problems that we would ask them to solve. This may not reflect that sometimes programmers are motivated to read the code by a concrete task at hand, such as fixing a specific bug. In terms of problem-solving, we designed the first one similar to an open exam and the second one as a resemblance of a closed exam. The former permitted access to the internet and AI tools, whereas the latter did not. This major difference may have led to participants' weaker task performance in Task 2, compared to Task 1. However, such a design may have mitigated the potential overwhelming outcome for participants in Task 1 and positively scaffolded their engagement with the experiment by allowing time to ramp up.

\textbf{Internal Validity.}~Understanding of a non-trivial program and learning are difficult to measure. Although we use hands-on code changes to test participants' understanding, and intend to cover both front-end and back-end oriented aspects, it is just one way to measure their understanding. Participants' understanding of the programs may manifest in other tasks or aspects. In addition, some participants may be influenced by the monetary incentive, rendering some degree of underperformance or overperformance. Given that all participants receive the same amount of compensation, this effect is expected to be balanced between the two groups. Further, our measurement of the influence on learning experience primarily relies on participants' self-assessment data. However, it is challenging (if not impossible) for participants to explicitly separate the natural learning effect between the two tasks from the learning introduced by our tool design. 

\textbf{External Validity.}~Our study validated the influence of gaze-based assistance only on Java web applications. The gaze patterns and their impact on the measures may not appear the same in codebases of different programming languages and architectures. However, projects using object-oriented programming languages and MVC-like architectures shall generate similar findings. Additionally, our tool design and experiment were confined to the IDE. Programmers' reading behaviour and comprehension strategy may be different in other types of programming environments, such as a cloud-based web editor. This is because the layouts and features of other programming environments may be largely different, as well as their ways of organizing the code. Finally, our findings of how our tool influenced the reading patterns and cognitive load are restricted to novice programmers, who are students without much industrial experience. Other groups of programmers, for instance, professional software developers, may not demonstrate the same patterns or experience the same cognitive load while using our tool. 
\section{Conclusion}
\label{sec:conclusion}

To conclude, we conducted a controlled experiment alongside a pre-experiment survey and post-experiment interviews with 40 novice Java programmers who were divided into two groups for comparison. We examined whether our tool GazePrinter, which visualizes expert gaze for attention direction, assisted programmers in reducing the time spent comprehending an unseen codebase and the cognitive load experienced, achieving better task performance, and modulating their gaze behaviour and learning experience. While the experiment group exhibited closer alignment with the expert gaze in terms of reading order and file coverage, and reported slightly less cognitive load, they underperformed the control group in response time and task performance. None of the aforementioned differences is statistically significant except for their difference in the alignment of reading strategies with experts. Nevertheless, our work corroborates and extends knowledge characterizing novice programmers in program comprehension. We also bridge the knowledge gap by adopting an intermediate-sized codebase in an eye-tracking study. The design and implementation of our tool, GazePrinter, further exemplify alternative ways to approach gaze-orienting assistance for program comprehension in an IDE that programmers use regularly. For future work, we suggest that researchers investigate gaze-based assistance for other programmer groups and conduct studies in AI-driven programming environments and with programs mixing both code and natural language prompts. 
\begin{acks}
We thank Ningzhi Tang, the first author of CodeGRITS, for the fruitful discussions while we were planning this study. We thank Diederick C. Niehorster for his feedback on various versions of this manuscript. We also thank all the participants for taking part in this study and for sharing their experiences and insights related to program comprehension. 

We would further like to thank the following funders who partly funded this work: the Swedish Foundation for Strategic Research (grant nbr. FFL18-0231), the Swedish Research Council (grant nbr. 2019-05658), the Swedish strategic research environment ELLIIT, the Wallenberg AI, Autonomous Systems and Software Program (WASP) funded by the Knut and Alice Wallenberg Foundation, and Dieter Schwarz Foundation (DSS).
\end{acks}

%%
%% The next two lines define the bibliography style to be used, and
%% the bibliography file.
\bibliographystyle{ACM-Reference-Format}
\bibliography{ref.bib}

\end{document}